# Layout de Cabine, Densidade de Assentos e Segmentação de Passageiros no Transporte Aéreo: Implicações para Preços, Receitas Auxiliares e Eficiência

Alessandro V. M. Oliveira✈
Instituto Tecnológico de Aeronáutica - ITA
Moisés D. Vassallo
Universidade Federal de Itajubá - UNIFEI

✈Autor correspondente. Praça Marechal Eduardo Gomes, 50. São José dos Campos, SP - Brasil.
E-mail address: alessandro@ita.br.

***Resumo***: **O presente estudo investiga como o layout e a densidade de assentos na cabine de aeronaves influenciam a precificação das passagens aéreas em voos domésticos. A análise baseia-se em microdados de cartões de embarque vinculados a entrevistas presenciais com passageiros, permitindo relacionar o preço pago à localização no mapa de assentos da aeronave, bem como a características de mercado e de operações de voo. Foram estimados modelos econométricos com uso do procedimento Post-Double-Selection LASSO (PDS-LASSO) de forma a selecionar inúmeros controles de fatores não observáveis atrelados a aspectos comerciais e operacionais, e assim possibilitar uma melhor identificação do efeito de variáveis como antecedência de compra, motivo da viagem, preço do combustível, estrutura de mercado, taxa de ocupação, dentre outras. Os resultados permitem inferir que uma maior densidade de fileiras de assentos está associada a preços menores, refletindo economias com o incremento do tamanho da aeronave e ganhos de eficiência operacional. Também foi obtido o resultado inesperado de que, em situações onde não havia cobrança de tarifa de seleção de assentos, passageiros com passagens mais caras eram frequentemente alocados em assentos do meio, devido à compra com baixa antecedência, quando as alternativas laterais já não estavam disponíveis. Esse comportamento contribui para explicar a lógica econômica de uma das principais receitas auxiliares das companhias aéreas. Além da análise quantitativa, o estudo incorpora uma abordagem exploratória sobre conceitos inovadores de cabine e seus possíveis efeitos sobre densidade e conforto a bordo.**

*Palavras-chave*: densidade de assentos, layout de cabine, precificação aérea, economias de densidade, eficiência operacional, gestão de receita, modelagem LASSO, inferência em alta dimensão.

## I. INTRODUÇÃO

A configuração interna das aeronaves comerciais, em especial a disposição e a densidade dos assentos, tem papel determinante na estrutura de custos e na percepção de valor dos serviços aéreos. A decisão sobre quantas fileiras instalar e quanto espaço oferecer entre poltronas envolve um equilíbrio entre eficiência operacional e conforto percebido dos passageiros, elementos que influenciam diretamente a formação dos custos operacionais, os preços das passagens e a segmentação de mercado das companhias aéreas.

No contexto da aviação doméstica brasileira, dominada pela operação de aeronaves narrowbody em classe econômica, as empresas passaram a adotar, a partir da última década, estratégias de diferenciação de produto baseadas no layout da cabine. Esse movimento incluiu a oferta de versões de econômica premium caracterizadas sobretudo pelo espaço adicional entre poltronas, além da ampliação da cobrança pela marcação antecipada de assentos. Esse movimento difere do segmento internacional, onde práticas de segmentação e distinção de classes já eram mais consolidadas desde períodos anteriores. Essas transformações ocorreram simultaneamente à disseminação de modelos de negócios orientados à eficiência de custos, em que o aumento da densidade de assentos é visto como um mecanismo eficaz de redução do custo por assento-quilômetro (CASK). Contudo, os efeitos quantitativos dessa densidade sobre os preços efetivamente pagos pelos passageiros ainda são pouco conhecidos, sobretudo em mercados domésticos com estruturas concorrenciais concentradas e forte heterogeneidade de perfis de viajantes.

Este estudo analisa a relação entre layout de cabine, densidade de assentos e preços das passagens aéreas no mercado doméstico brasileiro. A análise utiliza uma base inédita construída a partir da vinculação de entrevistas presenciais com passageiros e microdados de cartões de embarque coletados em 2014. Essa combinação permite identificar o valor efetivamente pago por cada viajante e relacioná-lo de forma precisa às características do voo, do assento, da configuração interna da aeronave e do operador. O estudo também desenvolve uma discussão qualitativa sobre a forma como o layout de cabine influencia conforto, percepção de valor e estratégias comerciais das companhias aéreas. Essa discussão aborda inovações de design propostas ao longo dos últimos anos por fabricantes e desenvolvedores, incluindo os projetos Morph Seating, Cozy Suite, Air Lair, StepSeat, FlexSeat, Zephyr Seat, Cabin Hexagon, Checkerboard e Skyrider. Cada uma dessas propostas envolve mudanças estruturais na utilização do espaço interno e pode afetar custos operacionais, densidade da cabine, ergonomia e possíveis variações na disposição a pagar dos passageiros. Assim, o trabalho combina análise econométrica e reflexão conceitual para examinar o papel da configuração física das aeronaves na formação de preços.

A base de dados empregada contém informações detalhadas sobre o perfil de compra do passageiro. Ao vincular esses elementos ao assento ocupado pelo viajante, torna-se possível observar se diferenças estruturais da cabine e da localização do passageiro apresentam associação estatística com o preço pago. Para explorar essa relação utilizou-se o método econométrico Post-Double-Selection LASSO (PDS-LASSO). Esse método realiza seleção de variáveis em ambientes de alta dimensão e permite reduzir parte do viés de variável omitida ao identificar subconjuntos de controles relacionados tanto ao preço quanto às variáveis de interesse. O conjunto inicial de controles incluía efeitos de data da pesquisa, horário do voo, aeroportos de origem e destino e características individuais do passageiro. Esses grupos foram submetidos ao procedimento de seleção que extrai controles estatisticamente relevantes para cada especificação. Essa abordagem contribui para melhorar a precisão das estimativas, embora mantenha limites inerentes a métodos de regularização e à estrutura própria dos dados utilizados.

O trabalho está estruturado da seguinte forma. Após esta introdução, a Seção II apresenta a revisão da literatura sobre layout de cabine, conforto e efeitos de mercado. A Seção III discute o caso de estudo, incluindo a evolução histórica das configurações no Brasil, a descrição de conceitos inovadores existentes na indústria e uma visualização de caráter exploratório de um layout de cabine de alta densidade. Na sequência, o arcabouço conceitual que dá suporte à modelagem empírica é discutido na Seção IV, seguido pela apresentação dos dados na Seção V e pelos procedimentos metodológicos e econométricos na Seção VI. A Seção VII expõe os resultados das estimações e, por fim, a Seção VIII sintetiza as conclusões do estudo.

## II. LAYOUT DE CABINE, CONFORTO E EFEITOS DE MERCADO

A literatura sobre os efeitos da configuração interna de cabine em relação aos assentos das aeronaves não é abundante. Em geral, pode ser dividida nos seguintes temas: I) nível de conforto percebido pelo passageiro; II) inovações e práticas comerciais das companhias aéreas; e III) disposição a pagar por conforto e diferenciação de produto aéreo. A seguir, apresenta-se uma revisão dessa literatura.

### II.1. NÍVEL DE CONFORTO PERCEBIDO PELO PASSAGEIRO

Kremser et al. (2012), Miller, Lapp & Parkinson (2019) e Anjani et al. (2020) investigam a relação entre o pitch do assento[1] e o conforto a bordo percebido pelo passageiro. Miller, Lapp & Parkinson (2019) incluem a taxa de ocupação e dados demográficos dos passageiros em sua análise dessa relação. Anjani et al. (2020) contribuem analisando outros fatores influenciadores, como a experiência espacial e as medidas antropométricas. Esses autores efetuaram uma pesquisa junto a duzentos e noventa e quatro participantes, que experimentaram assentos de classe econômica em um Boeing 737 com pitches variados de assento, entre 28 e 34 polegadas. Na pesquisa, foram coletadas medidas antropométricas dos participantes. Os autores fizeram perguntas aos respondentes acerca dos níveis de conforto e desconforto percebidos. Para definir "experiência espacial" os pesquisadores apresentaram afirmações para serem avaliadas pelos respondentes, do tipo "*Eu me sinto restringido pela distância das fileiras de assentos*" e "*Sinto-me estressado por causa da distância das fileiras de assentos*", dentre outras. O estudo encontrou uma relação significativa entre o pitch do assento e as medidas de conforto e desconforto do passageiro. Além disso, obteve-se evidências de que a classificação média dada pelo entrevistado ao nível desconforto de cada pitch com relação ao assento do meio era sistematicamente maior do que a janela e o assento do corredor. Atestaram, entretanto, que o pitch do assento aparentemente afeta o (des)conforto dos passageiros mais do que a localização do mesmo na cabine. Por fim, obtiveram evidências de que as medidas antropométricas afetam significativamente o (des)conforto em tamanhos de pitch de assento menores.

---

[1] O "pitch" do assento é a distância entre um ponto qualquer de um assento e o mesmo ponto do assento na mesma posição em fileira imediatamente anterior ou posterior, medido em centímetros ou polegadas. Em geral, os pitches dos assentos da classe econômica variam entre 28 e 34 polegadas, ou entre 71 a 86 cm (com arredondamento, dado que uma polegada equivale a 2,54 cm).

No Brasil, a temática do conforto dos passageiros e da relação com o perfil antropométrico da população brasileira, bem como os projetos de poltronas aeronáuticas, foi abordado por uma sequência de estudos entre o final dos anos 2000 e início dos anos 2010. Dentre esses estudos, destacam-se Silva & Monteiro (2009), Souza (2010), Rossi (2011), Greghi (2012) e Silva Filho, Andrade & Ciaccia (2012).

## II.2. Inovações e práticas comerciais de companhias aéreas

Rothkopf & Wald (2011) estudam a introdução de inovações no serviço provido pelas companhias aéreas, como entretenimento a bordo com telas de led individuais nas poltronas da classe econômica, o web check-in, os totens de atendimento nos aeroportos e o uso de smartfones a bordo. Algumas das inovações relacionadas ao layout interno das aeronaves abordadas são a política de bloqueio do assento do meio em classes executivas em voos de curta etapa em aeronaves de fuselagem estreita e a utilização de "cabine híbrida", onde há uma distinção de serviços dentro da classe econômica, como por exemplo a criação de classe econômica premium com pitch de assento maior. Os autores analisam uma amostra de trinta companhias aéreas e detectam padrões de inovações utilizadas pelas mesmas, dentro de diversas categorias de inovação identificadas. Os autores sugerem a existência de prioridades diferentes de inovação para os diferentes modelos de negócios das companhias aéreas existentes.

A maior parte dos estudos da literatura relacionados à questão econômica do design de assentos na aviação comercial centra-se na questão da precificação (tarifas) da marcação de assento das companhias aéreas e suas consequências. A seguir, apresentamos alguns desses estudos.

Kyparisis & Koulamas (2018) estudam o preço e a alocação ideal de assentos para um problema de gerenciamento de receitas de companhias aéreas, considerando a existência de duas cabines, no qual há uma divisória flexível das cabines de negócios e econômica. Os autores identificam a divisão ideal da cabine e as tarifas ideais para ambas as cabines com uma formulação matemática geral, com base em uma função multiplicativa de preço-demanda. Consideram três distribuições aleatórias diferentes de demanda por viagens. Adicionalmente, apresentam evidências de que o particionamento ideal e os preços ideais não são sensíveis às distribuições aleatórias de demanda. O estudo também aborda o efeito de uma restrição de capacidade na cabine de negócios, e concluem que essa restrição produziria o efeito de elevar as tarifas da classe executiva e geraria queda de receita total.

Mumbower, Garrow & Newman (2015) estudam as compras de serviços de seleção de assentos na classe econômica premium e suas implicações para estratégias de preços ideais das companhias aéreas. Os autores utilizam um banco de dados de preços e de exibições online de mapas de assentos, coletadas no site da companhia aérea norte-americana JetBlue Airways. Os resultados do estudam apontam que múltiplos fatores influenciam o comportamento de compra online, incluindo a taxa de marcação de assentos, a antecedência de compra, o número de passageiros viajando juntos e os fatores de aproveitamento das aeronaves - observados online, na medida em que o mapa de assentos do voo era exibido. O estudo proporciona evidências de que os clientes têm entre 2 e 3,3 vezes mais chances de comprar assentos de classe econômica premium - configurada com espaço extra para as pernas e privilégios de embarque antecipado -, quando em toda a aeronave não restam assentos na janela ou corredor que possam ser adquiridos gratuitamente. Os resultados dos autores ainda sugerem que os clientes que compram bilhetes mais próximos à data de partida são menos sensíveis ao preço e estão dispostos a pagar taxas mais altas de marcação de assentos. Interessantemente, os autores ainda inferem, a partir de seus resultados, que as taxas de assentos da JetBlue estariam subfaturadas em muitos mercados, e que uma tarifa de marcação estática ideal aumentaria as receitas em 8%, enquanto que as taxas dinâmicas ideais - aquelas que evoluem ao longo do período de reservas - aumentariam as receitas em 10,2%. Por fim, sugerem uma política corporativa que faria a empresa aumentar potencialmente suas receitas em 12,8%, por meio do uso combinado de uma política de manutenção das taxas de assentos aos níveis correntes, em conjunto com a reserva de certas fileiras de assentos para clientes premium, como fizeram posteriormente várias companhias aéreas, inclusive brasileiras.

Rouncivell, Timmis & Ison (2018) estudam a disposição a pagar pela marcação de assento em voos domésticos no Reino Unido, utilizando um método de preferência declarada. Os autores investigam a relação entre as características e opiniões dos passageiros, com valores por eles declarados para pagamento de taxas de marcação de assentos. Apontam os autores que a sensibilidade dos consumidores aos preços das passagens, tanto para viagens business quanto para não-business (lazer e outros motivos pessoais), encontra-se negativamente correlacionada com a disposição declarada a pagar pela marcação de assento. Por outro lado, as percepções dos clientes quanto à reputação da companhia aérea e à conveniência dos horários de seus voos estão positivamente correlacionadas com a disposição a pagar pela marcação de assentos pelos viajantes não-business. Além disso, a compra anterior de um produto de seleção de assentos está fortemente correlacionada com a disposição futura a pagar pela seleção de assentos, para ambos os segmentos de passageiros. Esse resultado é interpretado pelos autores como sendo advindo do fato de, nessa situação, os consumidores serem mais capazes de valorizar os benefícios de seu assento escolhido a partir da experiência passada acumulada.

Shao, Kauermann & Smith (2020) estudam a decisão de adquirir a seleção de assentos e, dada essa decisão, quando e quais assentos são selecionados pelos passageiros. Para responder a essas perguntas, utilizam um conjunto de dados de 485.279 reservas em cinco rotas intercontinentais, extraídas do histórico de reservas de uma grande companhia aérea europeia. Os autores encontram o que reputam como fortes evidências de comportamentos de compra dos passageiros de "evitar assentos do meio" e "preferir assentos dianteiros". Seus resultados também sugerem que a probabilidade de adquirir a marcação de assentos depende do seu preço em relação ao preço da passagem, do canal de distribuição de bilhetes utilizado, da antecedência de compra e de efeitos sazonais.

Zhou et al. (2020) estudam a disposição a pagar pela seleção de assentos na classe econômica na aviação chinesa. Utilizam uma abordagem de métodos mistos, combinando entrevistas individuais com uma pesquisa online, de forma a explorar fatores que influenciam a disposição dos consumidores aéreos a pagar pela marcação de assentos na classe econômica. Encontram os autores que tanto fatores como intrínsecos, como a duração da viagem, conforto dos assentos e a conveniência, quanto extrínsecos, como situações de pagamento e consumo, têm um impacto significativo na disposição dos consumidores chineses a pagar pela marcação de assentos.

Por fim, uma das propostas mais radicais em relação à modificação do layout de cabines aéreas é a ideia de transportar passageiros em pé. Um exemplo concreto dessa abordagem é o Skyrider, cujo conceito será descrito na próxima seção. Um dos poucos estudos que analisam essa possibilidade é Romli et al. (2014), que investigam a viabilidade de reduzir custos operacionais e, consequentemente, os preços das passagens por meio do aumento da capacidade de passageiros por voo. O estudo parte do princípio de que essa expansão poderia diluir custos fixos e tornar o transporte aéreo mais acessível, especialmente em mercados sensíveis ao preço, como o de viagens curtas e de baixo custo. Com base em um estudo de caso no mercado malaio, os autores concluem que, embora teoricamente viável em voos curtos, sua adoção dependeria de análises adicionais sobre segurança, conforto e impactos estruturais nas aeronaves.

## II.3. DISPOSIÇÃO A PAGAR POR CONFORTO E DIFERENCIAÇÃO DE PRODUTO AÉREO

O estudo de Lee & Luengo-Prado (2004) é um dos trabalhos centrais sobre o valor econômico do conforto a bordo, em particular referente ao espaço entre poltronas. Os autores investigam se passageiros pagaram tarifas relativamente mais altas após duas iniciativas implementadas em 2000 por grandes companhias aéreas norte-americanas: o programa "More Room Throughout Coach", da American Airlines, e o "Economy Plus", da United Airlines. Ambas ampliaram o espaço entre fileiras na classe econômica, mas com estratégias distintas. A American aumentou o pitch em todos os assentos, enquanto a United aplicou o aumento apenas a algumas fileiras, direcionadas sobretudo a passageiros corporativos e clientes de alta fidelidade.

A análise utiliza dados de painel do OD1A, cobrindo 1998 a 2002 e aproximadamente mil pares de aeroportos com sobreposição de rotas entre as "Big Six". A variável dependente é a tarifa média de ida e volta. Entre as variáveis de controle estão participação de mercado, distância do itinerário, participação no aeroporto de origem, frequência de voos, proxy de passageiros corporativos, oferta de voo direto, pontualidade, alavancagem financeira e tendência temporal. Esses controles reduzem a influência de fatores estruturais que diferenciam as companhias, embora o efeito do pitch seja capturado indiretamente por meio de dummies pré e pós para American e United, dado que a base de dados dos autores não contém medidas explícitas de configuração interna.

Os modelos de efeitos fixos por par de aeroportos dos autores sugerem resultados alinhados à literatura de diferenciação de produtos. No caso da American Airlines, o aumento generalizado do espaço não se traduziu em prêmios tarifários. Os coeficientes estimados indicam queda do prêmio relativo da American após a intervenção e redução adicional de suas tarifas em comparação ao próprio período pré. Já para a United Airlines, observou-se aumento do prêmio relativo após a adoção do Economy Plus, da ordem de US$ 11 no modelo agregado, sugerindo que a estratégia segmentada direcionada a passageiros menos sensíveis ao preço foi mais eficaz na geração de diferenciação percebida.

Os resultados evidenciam que a disposição a pagar por conforto é heterogênea. Passageiros a lazer tendem a priorizar preço, enquanto passageiros corporativos tendem a valorizar mais espaço e conveniência. Assim, o desempenho superior da United está associado ao foco em segmentos de maior yield, consistente com modelos de competição espacial e segmentação por elasticidade. O artigo ressalta que parte das diferenças encontradas pode resultar de outros atributos de serviço não mensurados e que o impacto total em receita depende também de possíveis ganhos de ocupação. Isso sugere a necessidade de pesquisas que incorporem variáveis operacionais, comportamentais e medidas mais diretas de configuração interna.

Em síntese, Lee & Luengo-Prado (2004) oferecem evidência de que a ampliação do espaço entre assentos pode gerar diferenciação tarifária quando implementada de forma segmentada e orientada a passageiros de alto valor. O estudo reforça a relevância da heterogeneidade das preferências e fornece base empírica e metodológica importante para análises atuais sobre layout de cabine, conforto e precificação.

No presente trabalho, conforme veremos, o objetivo de pesquisa e a abordagem quantitativa é equivalente à utilizada por Lee & Luengo-Prado (2004), no sentido de que também buscamos identificar diferenças de precificação de passagens aéreas associadas à configuração interna da cabine. A principal distinção reside no fato de que dispomos de microdados com a localização individual do assento, o que amplia o escopo analítico e permite examinar efeitos distributivos dentro da própria aeronave. Por outro lado, não contamos com um experimento temporal do tipo antes e depois, como no caso das intervenções da American e da United. Nossa identificação decorre da comparação entre aeronaves que adotaram e que não adotaram determinadas configurações, conforme será detalhado adiante. O objetivo permanece essencialmente o mesmo, avaliar se alterações no layout a bordo geram diferenças sistemáticas nos preços pagos pelos passageiros.

## III. Caso de estudo

Esta seção apresenta o caso de estudo utilizado no trabalho, baseado no transporte aéreo doméstico brasileiro. Primeiro, descreve-se a evolução do layout de cabine adotado por empresas que operam e operavam no país, destacando mudanças estruturais relevantes para a densidade de assentos e para o espaço disponível ao passageiro. Em seguida, são introduzidos conceitos e propostas recentes de configuração de cabine desenvolvidos pela indústria, que ajudam a contextualizar alternativas de uso do espaço e potenciais implicações para conforto e custos. Por fim, realiza-se uma inspeção exploratória de arranjos alternativos aplicados a aeronaves típicas da aviação nacional, com o objetivo de ilustrar como diferentes escolhas de layout podem afetar a organização interna da cabine. Essa abordagem fornece o pano de fundo necessário para o modelo conceitual e para a análise econométrica apresentados nas seções seguintes.

### III.1. Evolução da configuração de layout de cabine no Brasil

O transporte aéreo no Brasil passou por inúmeras transformações ao longo das últimas décadas. Com a desregulação econômica dos anos 1990 e início dos anos 2000, as companhias aéreas obtiveram liberdade estratégica na formação de preços, frequências de voo e malha aérea. Permitiu-se a livre mobilidade das operadoras, de forma que pudessem iniciar ou encerrar voos com grande agilidade. A entrada de novas empresas foi facilitada, restringindo-se aos critérios de certificação técnica. Os dois maiores eventos ocorridos desde então nesse setor foram as entradas das empresas Gol e Azul em 2001 e 2008, respectivamente. Em ambos os casos, houve um acirramento da concorrência nos primeiros anos pós-entrada, fruto da precificação de penetração de mercado das novatas, causando grande otimismo quanto às possibilidades de crescimento do setor.

Mais importante do que o efeito de curto prazo da nova concorrência foram as estratégias de posicionamento de mercado observadas. A Gol iniciou suas operações fortemente baseada no modelo de negócios de baixo custo inspirado na norte-americana Southwest Airlines, e, mesmo com altos e baixos em seus reposicionamentos de mercado, logrou manter essa diretriz ao longo do tempo. A presença de rival "low cost" levou à incumbente TAM - atual Latam - a uma concorrência em custos, onde a exploração das economias de densidade de tráfego por meio de aeronaves maiores e da densificação das cabines com fileiras de assentos era a tônica. Igualmente, com a entrada da Azul, houve forte concorrência em preços, muito embora a empresa entrante operasse inicialmente a partir do aeroporto secundário de Campinas/Viracopos, e com operações de jatos narrowbody de menor capacidade de assentos da Embraer.

Passados alguns anos da década de 2010, observou-se no setor o surgimento de um novo paradigma para a concorrência entre as transportadoras existentes: a corrida pela geração de receitas auxiliares, com cobranças de taxas extras e tarifas customizadas em "pacotes" de atributos. Esse movimento, que teve seu ápice com a introdução de tarifas de despacho de bagagem em 2017, iniciou-se antes, com a ruptura com as estratégias puramente voltadas à densificação das cabines das aeronaves, a partir da introdução dos espaços extras nas primeiras fileiras voltados aos passageiros com disposição a pagar por essa amenidade a bordo.

Em novembro de 2013, a Gol introduziu nos voos da ponte aérea Rio de Janeiro-São Paulo (par de aeroportos Congonhas-Santos Dumont) o seu assento "GOL+Conforto". A inovação consistia na oferta, mediante cobrança de taxa extra, em seus Boeing 737-800NG, marcação de assento em poltronas com maior pitch, de 34 polegadas, face às 30 polegadas anteriormente em vigor, com reclinação 50% maior e assento do meio bloqueado. Esses atributos estavam presentes até a sétima fileira das aeronaves, que passaram a ter capacidade de 177 assentos[2]. A partir do final de 2013, a empresa ampliou sua operação de aeronaves configurada com assentos GOL+Conforto, atingindo a 100% da malha doméstica em outubro de 2014[3].

A sequência de figuras a seguir apresenta mapas ilustrativos das configurações de layout das aeronaves em operação no período analisado. Os diagramas não estão em escala real e utilizam simbologia própria adotada pelos autores, detalhada no rodapé de cada imagem, indicando componentes essenciais da cabine, como galley, banheiros, asas, saídas de emergência, última fileira e posições específicas de assentos. Atenção particular deve ser dada às células marcadas com o número 1, que podem aparecer com ou sem sublinhado, sendo que o sublinhado identifica a existência de um assento com pitch extra oferecido pela companhia aérea. Esses mapas, construídos a partir de análises dos mapas do Guia Panrotas em edições históricas, têm caráter representativo e servem para visualizar diferenças estruturais relevantes entre aeronaves e companhias aéreas no momento em que o produto GOL+Conforto foi introduzido.

A Figura 1 apresenta duas configurações possíveis do Boeing 737-800 da Gol, aeronave fabricada pela Boeing e amplamente utilizada no mercado doméstico brasileiro. À esquerda, mostra-se o layout com 177 assentos, no qual as sete primeiras fileiras possuem poltronas com espaço extra, indicadas pelo "1" sublinhado. À direita, apresenta-se a versão com 178 assentos, sem qualquer zona de maior espaço, refletindo a configuração padrão de alta densidade adotada em parte da frota no período analisado.

Na sequência, as Figuras 2, 3 e 4 ilustram exemplos de mapas de assentos de aeronaves das principais companhias aéreas à época da introdução do assento GOL+Conforto. Na Figura 2, observa-se à esquerda o ATR-72-600 da Azul, fabricado pela ATR, com 68 assentos, e à direita o Embraer E-195, com 118 assentos. A Figura 3 apresenta, à esquerda, o Airbus A319 da Avianca Brasil, com 132 assentos, e, à direita, o Airbus A320, também da Avianca Brasil, com 162 assentos. Por fim, a Figura 4 mostra dois layouts do Airbus A320 operado pela TAM, atual Latam, sendo a aeronave à esquerda configurada com 156 assentos e a da direita configurada com 174 assentos.

---

[2] Fonte: www.melhoresdestinos.com.br.
[3] Fonte: www.passageirodeprimeira.com.

| Boeing 737-800 (Gol) 177 SEATS | | | | | | | | | | Boeing 737-800 (Gol) 178 SEATS | | | | | | | | | |
|---|---|---|---|---|---|---|---|---|---|---|---|---|---|---|---|---|---|---|---|
| row | ‡ | A | B | C | D | E | F | ‡ | | row | ‡ | A | B | C | D | E | F | ‡ | |
| | | | W | | | G | | | | | | | W | | | G | | | |
| 1 | | 1 | 1 | 1 | 1 | 1 | 1 | | | 1 | | 1 | 1 | 1 | 1 | 1 | 1 | | |
| 2 | | 1 | 1 | 1 | 1 | 1 | 1 | | | 2 | | 1 | 1 | 1 | 1 | 1 | 1 | | |
| 3 | | 1 | 1 | 1 | 1 | 1 | 1 | | | 3 | | 1 | 1 | 1 | 1 | 1 | 1 | | |
| 4 | | 1 | 1 | 1 | 1 | 1 | 1 | | | 4 | | 1 | 1 | 1 | 1 | 1 | 1 | | |
| 5 | | 1 | 1 | 1 | 1 | 1 | 1 | | | 5 | | 1 | 1 | 1 | 1 | 1 | 1 | | |
| 6 | | 1 | 1 | 1 | 1 | 1 | 1 | | | 6 | | 1 | 1 | 1 | 1 | 1 | 1 | | |
| 7 | | 1 | 1 | 1 | 1 | 1 | 1 | | | 7 | | 1 | 1 | 1 | 1 | 1 | 1 | | |
| 8 | | 1 | 1 | 1 | 1 | 1 | 1 | | | 8 | | 1 | 1 | 1 | 1 | 1 | 1 | | |
| 9 | | 1 | 1 | 1 | 1 | 1 | 1 | | | 9 | | 1 | 1 | 1 | 1 | 1 | 1 | | |
| 10 | / | 1 | 1 | 1 | 1 | 1 | 1 | | \ | 10 | / | 1 | 1 | 1 | 1 | 1 | 1 | | \ |
| 11 | / | 1 | 1 | 1 | 1 | 1 | 1 | | \ | 11 | / | 1 | 1 | 1 | 1 | 1 | 1 | | \ |
| 12 | / | 1 | 1 | 1 | 1 | 1 | 1 | | \ | 12 | / | 1 | 1 | 1 | 1 | 1 | 1 | | \ |
| 13 | / | 0 | 0 | 0 | 0 | 0 | 0 | | \ | 13 | / | 0 | 0 | 0 | 0 | 0 | 0 | | \ |
| 14 | / | 1 | 1 | 1 | 1 | 1 | 1 | | \ | 14 | / | 1 | 1 | 1 | 1 | 1 | 1 | | \ |
| 15 | / | 1 | 1 | 1 | 1 | 1 | 1 | | \ | 15 | / | 1 | 1 | 1 | 1 | 1 | 1 | | \ |
| 16 | / ‡ | 0 | 0 | 0 | 0 | 0 | 0 | ‡ | \ | 16 | / ‡ | 0 | 1 | 1 | 1 | 1 | 0 | ‡ | \ |
| 17 | / ‡ | 1 | 1 | 1 | 1 | 1 | 1 | ‡ | \ | 17 | / | 1 | 1 | 1 | 1 | 1 | 1 | | \ |
| 18 | / | 1 | 1 | 1 | 1 | 1 | 1 | | \ | 18 | / | 1 | 1 | 1 | 1 | 1 | 1 | | \ |
| 19 | / | 1 | 1 | 1 | 1 | 1 | 1 | | \ | 19 | / | 1 | 1 | 1 | 1 | 1 | 1 | | \ |
| 20 | / | 1 | 1 | 1 | 1 | 1 | 1 | | \ | 20 | / | 1 | 1 | 1 | 1 | 1 | 1 | | \ |
| 21 | | 1 | 1 | 1 | 1 | 1 | 1 | | | 21 | | 1 | 1 | 1 | 1 | 1 | 1 | | |
| 22 | | 1 | 1 | 1 | 1 | 1 | 1 | | | 22 | | 1 | 1 | 1 | 1 | 1 | 1 | | |
| 23 | | 1 | 1 | 1 | 1 | 1 | 1 | | | 23 | | 1 | 1 | 1 | 1 | 1 | 1 | | |
| 24 | | 1 | 1 | 1 | 1 | 1 | 1 | | | 24 | | 1 | 1 | 1 | 1 | 1 | 1 | | |
| 25 | | 1 | 1 | 1 | 1 | 1 | 1 | | | 25 | | 1 | 1 | 1 | 1 | 1 | 1 | | |
| 26 | | 1 | 1 | 1 | 1 | 1 | 1 | | | 26 | | 1 | 1 | 1 | 1 | 1 | 1 | | |
| 27 | | 1 | 1 | 1 | 1 | 1 | 1 | | | 27 | | 1 | 1 | 1 | 1 | 1 | 1 | | |
| 28 | | 1 | 1 | 1 | 1 | 1 | 1 | | | 28 | | 1 | 1 | 1 | 1 | 1 | 1 | | |
| 29 | | 1 | 1 | 1 | 1 | 1 | 1 | | | 29 | | 1 | 1 | 1 | 1 | 1 | 1 | | |
| 30 | | 1 | 1 | 1 | 1 | 1 | 1 | | | 30 | | 1 | 1 | 1 | 1 | 1 | 1 | | |
| 31 | | 1 | 1 | 1 | 1 | 1 | 1 | | | 31 | | 1 | 1 | 1 | 1 | 1 | 1 | | |
| 32 | | 1 | 1 | 1 | | | | | | | | | W | | | | W | | |
| | | | W | | | W | | | | ‡ | | | | G | | | | ‡ | |
| | ‡ | | | G | | | | ‡ | | | | | | | | | | | |

*Símbolos: "G" (galley); "W" (banheiro); "/" e "\" (asa); "‡" saída de emergência; "A", "B", "C" e "D" (posição do assento); "row" (fileira); "0" (assento não existente); "1" assento existente"; "1" assento existente com maior pitch". Notas: mapas não estão em escala real; tamanhos dos retângulos indicativos de assentos não refletem as proporções de pitch e largura dos mesmos; posições aproximadas dos componentes da cabine da aeronave; uma fileira contendo apenas "0" indica inexistência do espaço da mesma. Fontes: Guia PANROTAS 2014, n. 490-493 e website www.web.archive.org/web/www.seatguru.com (ano 2014).*

**Figura 1 - Mapa de assentos - aeronaves Gol (2014) - exemplos**

| ATR-72-600 (Azul) | | | | | |
|---|---|---|---|---|---|
| 68 SEATS | | | | | |
| row | | G | | G | |
| | ‡ | A B | | C D | ‡ |
| 1 | | 1 1 | | 1 1 | |
| 2 | | 1 1 | | 1 1 | |
| 3 | | 1 1 | | 1 1 | |
| 4 | | 1 1 | | 1 1 | |
| 5 | | 1 1 | | 1 1 | |
| 6 | / | 1 1 | | 1 1 | \ |
| 7 | / | 1 1 | | 1 1 | \ |
| 8 | / | 1 1 | | 1 1 | \ |
| 9 | / | 1 1 | | 1 1 | \ |
| 10 | / | 1 1 | | 1 1 | \ |
| 11 | / | 1 1 | | 1 1 | \ |
| 12 | / | 1 1 | | 1 1 | \ |
| 13 | | 1 1 | | 1 1 | |
| 14 | | 1 1 | | 1 1 | |
| 15 | | 1 1 | | 1 1 | |
| 16 | | 1 1 | | 1 1 | |
| 17 | | 1 1 | | 1 1 | |
| | ‡ | | | | ‡ |
| | | W | | G | |

| Embraer E-195 (Azul) | | | | | |
|---|---|---|---|---|---|
| 118 SEATS | | | | | |
| row | | W | | G | |
| | ‡ | A B | | C D | ‡ |
| 1 | | 0 0 | | 1 1 | |
| 2 | | 1 1 | | 1 1 | |
| 3 | | 1 1 | | 1 1 | |
| 4 | | 1 1 | | 1 1 | |
| 5 | | 1 1 | | 1 1 | |
| 6 | | 1 1 | | 1 1 | |
| 7 | | 1 1 | | 1 1 | |
| 8 | | 1 1 | | 1 1 | |
| 9 | | 1 1 | | 1 1 | |
| 10 | / | 1 1 | | 1 1 | \ |
| 11 | / | 1 1 | | 1 1 | \ |
| 12 | / | 1 1 | | 1 1 | \ |
| 13 | / | 1 1 | | 1 1 | \ |
| 14 | / ‡ | 1 1 | | 1 1 | ‡ \ |
| 15 | / | 1 1 | | 1 1 | \ |
| 16 | / | 1 1 | | 1 1 | \ |
| 17 | / | 1 1 | | 1 1 | \ |
| 18 | / | 1 1 | | 1 1 | \ |
| 19 | / | 1 1 | | 1 1 | \ |
| 20 | / | 1 1 | | 1 1 | \ |
| 21 | | 1 1 | | 1 1 | |
| 22 | | 1 1 | | 1 1 | |
| 23 | | 1 1 | | 1 1 | |
| 24 | | 1 1 | | 1 1 | |
| 25 | | 1 1 | | 1 1 | |
| 26 | | 1 1 | | 1 1 | |
| 27 | | 1 1 | | 1 1 | |
| 28 | | 1 1 | | 1 1 | |
| 29 | | 1 1 | | 1 1 | |
| 30 | | 1 1 | | 1 1 | |
| | ‡ | | | | ‡ |
| | | W | | G | |

*Símbolos: "G" (galley); "W" (banheiro); "/" e "\" (asa); "‡" saída de emergência; "A", "B", "C" e "D" (posição do assento); "row" (fileira); "0" (assento não existente); "1" assento existente"; "1" assento existente com maior pitch". Notas: mapas não estão em escala real; tamanhos dos retângulos indicativos de assentos não refletem as proporções de pitch e largura dos mesmos; posições aproximadas dos componentes da cabine da aeronave; uma fileira contendo apenas "0" indica inexistência do espaço da mesma. Fontes: Guia PANROTAS 2014, n. 490-493 e website www.web.archive.org/web/www.seatguru.com (ano 2014).*

**Figura 2 - Mapa de assentos - aeronaves Azul (2014) - exemplos**

## Airbus A319 (Avianca Brasil)
### 132 SEATS

| row |   |   | W |   |   |   | G |   |   |   |
|-----|---|---|---|---|---|---|---|---|---|---|
|     |   | ‡ | A | B | C | D | E | F | ‡ |   |
| 1   |   |   | 1 | 1 | 1 | 1 | 1 | 1 |   |   |
| 2   |   |   | 1 | 1 | 1 | 1 | 1 | 1 |   |   |
| 3   |   |   | 1 | 1 | 1 | 1 | 1 | 1 |   |   |
| 4   |   |   | 1 | 1 | 1 | 1 | 1 | 1 |   |   |
| 5   |   |   | 1 | 1 | 1 | 1 | 1 | 1 |   |   |
| 6   |   |   | 1 | 1 | 1 | 1 | 1 | 1 |   |   |
| 7   | / |   | 1 | 1 | 1 | 1 | 1 | 1 |   | \ |
| 8   | / |   | 1 | 1 | 1 | 1 | 1 | 1 |   | \ |
| 9   | / |   | 1 | 1 | 1 | 1 | 1 | 1 |   | \ |
| 10  | / | ‡ | 1 | 1 | 1 | 1 | 1 | 1 | ‡ | \ |
| 11  | / |   | 1 | 1 | 1 | 1 | 1 | 1 |   | \ |
| 12  | / |   | 1 | 1 | 1 | 1 | 1 | 1 |   | \ |
| 13  | / |   | 0 | 0 | 0 | 0 | 0 | 0 |   | \ |
| 14  | / |   | 1 | 1 | 1 | 1 | 1 | 1 |   | \ |
| 15  |   |   | 1 | 1 | 1 | 1 | 1 | 1 |   |   |
| 16  |   |   | 1 | 1 | 1 | 1 | 1 | 1 |   |   |
| 17  |   |   | 1 | 1 | 1 | 1 | 1 | 1 |   |   |
| 18  |   |   | 1 | 1 | 1 | 1 | 1 | 1 |   |   |
| 19  |   |   | 1 | 1 | 1 | 1 | 1 | 1 |   |   |
| 20  |   |   | 1 | 1 | 1 | 1 | 1 | 1 |   |   |
| 21  |   |   | 1 | 1 | 1 | 1 | 1 | 1 |   |   |
| 22  |   |   | 1 | 1 | 1 | 1 | 1 | 1 |   |   |
| 23  |   |   | 1 | 1 | 1 | 1 | 1 | 1 |   |   |
|     |   |   | W |   |   | W |   |   |   |   |
|     |   | ‡ |   |   |   |   |   |   | ‡ |   |
|     |   |   |   |   | G |   |   |   |   |   |

## Airbus A320 (Avianca Brasil)
### 162 SEATS

| row |   |   | W |   |   |   | G |   |   |   |
|-----|---|---|---|---|---|---|---|---|---|---|
|     |   | ‡ | A | B | C | D | E | F | ‡ |   |
| 1   |   |   | 1 | 1 | 1 | 1 | 1 | 1 |   |   |
| 2   |   |   | 1 | 1 | 1 | 1 | 1 | 1 |   |   |
| 3   |   |   | 1 | 1 | 1 | 1 | 1 | 1 |   |   |
| 4   |   |   | 1 | 1 | 1 | 1 | 1 | 1 |   |   |
| 5   |   |   | 1 | 1 | 1 | 1 | 1 | 1 |   |   |
| 6   |   |   | 1 | 1 | 1 | 1 | 1 | 1 |   |   |
| 7   |   |   | 1 | 1 | 1 | 1 | 1 | 1 |   |   |
| 8   |   |   | 1 | 1 | 1 | 1 | 1 | 1 |   |   |
| 9   | / |   | 1 | 1 | 1 | 1 | 1 | 1 |   | \ |
| 10  | / |   | 1 | 1 | 1 | 1 | 1 | 1 |   | \ |
| 11  | / | ‡ | 1 | 1 | 1 | 1 | 1 | 1 | ‡ | \ |
| 12  | / | ‡ | 1 | 1 | 1 | 1 | 1 | 1 | ‡ | \ |
| 13  | / |   | 0 | 0 | 0 | 0 | 0 | 0 |   | \ |
| 14  | / |   | 1 | 1 | 1 | 1 | 1 | 1 |   | \ |
| 15  | / |   | 1 | 1 | 1 | 1 | 1 | 1 |   | \ |
| 16  | / |   | 1 | 1 | 1 | 1 | 1 | 1 |   | \ |
| 17  | / |   | 1 | 1 | 1 | 1 | 1 | 1 |   | \ |
| 18  | / |   | 1 | 1 | 1 | 1 | 1 | 1 |   | \ |
| 19  |   |   | 1 | 1 | 1 | 1 | 1 | 1 |   |   |
| 20  |   |   | 1 | 1 | 1 | 1 | 1 | 1 |   |   |
| 21  |   |   | 1 | 1 | 1 | 1 | 1 | 1 |   |   |
| 22  |   |   | 1 | 1 | 1 | 1 | 1 | 1 |   |   |
| 23  |   |   | 1 | 1 | 1 | 1 | 1 | 1 |   |   |
| 24  |   |   | 1 | 1 | 1 | 1 | 1 | 1 |   |   |
| 25  |   |   | 1 | 1 | 1 | 1 | 1 | 1 |   |   |
| 26  |   |   | 1 | 1 | 1 | 1 | 1 | 1 |   |   |
| 27  |   |   | 1 | 1 | 1 | 1 | 1 | 1 |   |   |
| 28  |   |   | 1 | 1 | 1 | 1 | 1 | 1 |   |   |
|     |   |   | W |   |   | W |   |   |   |   |
|     |   | ‡ |   |   |   |   |   |   | ‡ |   |
|     |   |   |   |   | G |   |   |   |   |   |

*Símbolos: "G" (galley); "W" (banheiro); "/" e "\" (asa); "‡" saída de emergência; "A", "B", "C" e "D" (posição do assento); "row" (fileira); "0" (assento não existente); "1" assento existente". Notas: mapas não estão em escala real; tamanhos dos retângulos indicativos de assentos não refletem as proporções de pitch e largura dos mesmos; posições aproximadas dos componentes da cabine da aeronave; uma fileira contendo apenas "0" indica inexistência do espaço da mesma. Fontes: Guia PANROTAS 2014, n. 490-493 e website www.web.archive.org/web/www.seatguru.com (ano 2014).*

**Figura 3 - Mapa de assentos - aeronaves Avianca (2014) - exemplos**

| row | | A | B | C | | D | E | F | | |
|---|---|---|---|---|---|---|---|---|---|---|
| | | **W** | | | | **G** | | | | |
| | ‡ | A | B | C | | D | E | F | ‡ | |
| 1 | | <u>1</u> | <u>0</u> | <u>1</u> | | <u>1</u> | <u>0</u> | <u>1</u> | | |
| 2 | | <u>1</u> | <u>0</u> | <u>1</u> | | <u>1</u> | <u>0</u> | <u>1</u> | | |
| 3 | | <u>1</u> | <u>0</u> | <u>1</u> | | <u>1</u> | <u>0</u> | <u>1</u> | | |
| 4 | | 0 | 0 | 0 | | 0 | 0 | 0 | | |
| 5 | | 1 | 1 | 1 | | 1 | 1 | 1 | | |
| 6 | | 1 | 1 | 1 | | 1 | 1 | 1 | | |
| 7 | | 1 | 1 | 1 | | 1 | 1 | 1 | | |
| 8 | | 1 | 1 | 1 | | 1 | 1 | 1 | | |
| 9 | / | 1 | 1 | 1 | | 1 | 1 | 1 | | \ |
| 10 | / | 1 | 1 | 1 | | 1 | 1 | 1 | | \ |
| 11 | / | 1 | 1 | 1 | | 1 | 1 | 1 | | \ |
| 12 | / ‡ | 1 | 1 | 1 | | 1 | 1 | 1 | ‡ | \ |
| 13 | / | 1 | 1 | 1 | | 1 | 1 | 1 | | \ |
| 14 | / | 1 | 1 | 1 | | 1 | 1 | 1 | | \ |
| 15 | / | 1 | 1 | 1 | | 1 | 1 | 1 | | \ |
| 16 | / | 1 | 1 | 1 | | 1 | 1 | 1 | | \ |
| 17 | / | 1 | 1 | 1 | | 1 | 1 | 1 | | \ |
| 18 | / | 1 | 1 | 1 | | 1 | 1 | 1 | | \ |
| 19 | | 1 | 1 | 1 | | 1 | 1 | 1 | | |
| 20 | | 1 | 1 | 1 | | 1 | 1 | 1 | | |
| 21 | | 1 | 1 | 1 | | 1 | 1 | 1 | | |
| 22 | | 1 | 1 | 1 | | 1 | 1 | 1 | | |
| 23 | | 1 | 1 | 1 | | 1 | 1 | 1 | | |
| 24 | | 1 | 1 | 1 | | 1 | 1 | 1 | | |
| 25 | | 1 | 1 | 1 | | 1 | 1 | 1 | | |
| 26 | | 1 | 1 | 1 | | 1 | 1 | 1 | | |
| 27 | | 1 | 1 | 1 | | 1 | 1 | 1 | | |
| 28 | | 1 | 1 | 1 | | 1 | 1 | 1 | | |

Airbus A320 (TAM) — 156 SEATS. Below row 28: W ... W ; ‡ ... ‡ ; G.

Airbus A320 (TAM) — 174 SEATS:

| row | | A | B | C | | D | E | F | | |
|---|---|---|---|---|---|---|---|---|---|---|
| | ‡ | | | | | | | | ‡ | |
| 1 | | 1 | 1 | 1 | | 1 | 1 | 1 | | |
| 2 | | 1 | 1 | 1 | | 1 | 1 | 1 | | |
| 3 | | 1 | 1 | 1 | | 1 | 1 | 1 | | |
| 4 | | 1 | 1 | 1 | | 1 | 1 | 1 | | |
| 5 | | 1 | 1 | 1 | | 1 | 1 | 1 | | |
| 6 | | 1 | 1 | 1 | | 1 | 1 | 1 | | |
| 7 | | 1 | 1 | 1 | | 1 | 1 | 1 | | |
| 8 | | 1 | 1 | 1 | | 1 | 1 | 1 | | |
| 9 | | 1 | 1 | 1 | | 1 | 1 | 1 | | |
| 10 | / | 1 | 1 | 1 | | 1 | 1 | 1 | | \ |
| 11 | / | 1 | 1 | 1 | | 1 | 1 | 1 | | \ |
| 12 | / ‡ | 1 | 1 | 1 | | 1 | 1 | 1 | ‡ | \ |
| 13 | / | 1 | 1 | 1 | | 1 | 1 | 1 | | \ |
| 14 | / | 1 | 1 | 1 | | 1 | 1 | 1 | | \ |
| 15 | / | 1 | 1 | 1 | | 1 | 1 | 1 | | \ |
| 16 | / | 1 | 1 | 1 | | 1 | 1 | 1 | | \ |
| 17 | / | 1 | 1 | 1 | | 1 | 1 | 1 | | \ |
| 18 | / | 1 | 1 | 1 | | 1 | 1 | 1 | | \ |
| 19 | / | 1 | 1 | 1 | | 1 | 1 | 1 | | \ |
| 20 | / | 1 | 1 | 1 | | 1 | 1 | 1 | | \ |
| 21 | | 1 | 1 | 1 | | 1 | 1 | 1 | | |
| 22 | | 1 | 1 | 1 | | 1 | 1 | 1 | | |
| 23 | | 1 | 1 | 1 | | 1 | 1 | 1 | | |
| 24 | | 1 | 1 | 1 | | 1 | 1 | 1 | | |
| 25 | | 1 | 1 | 1 | | 1 | 1 | 1 | | |
| 26 | | 1 | 1 | 1 | | 1 | 1 | 1 | | |
| 27 | | 1 | 1 | 1 | | 1 | 1 | 1 | | |
| 28 | | 1 | 1 | 1 | | 1 | 1 | 1 | | |
| 29 | | 1 | 1 | 1 | | 1 | 1 | 1 | | |

Below row 29: W ... W ; ‡ ... ‡ ; G.

Símbolos: "G" (galley); "W" (banheiro); "/" e "\" (asa); "‡" saída de emergência; "A", "B", "C" e "D" (posição do assento); "row" (fileira); "0" (assento não existente); "1" assento existente"; "<u>1</u>" assento existente com maior pitch". Notas: mapas não estão em escala real; tamanhos dos retângulos indicativos de assentos não refletem as proporções de pitch e largura dos mesmos; posições aproximadas dos componentes da cabine da aeronave; uma fileira contendo apenas "0" indica inexistência do espaço da mesma. Fontes: Guia PANROTAS 2014, n. 490-493 e website www.web.archive.org/web/www.seatguru.com (ano 2014).

**Figura 4 - Mapa de assentos - aeronaves TAM (2014) - exemplos**

## III.2. Conceitos inovadores de layout de cabine

O modelo de negócios low cost, associado a companhias aéreas como a Southwest Airlines nos Estados Unidos, e a Ryanair no Reino Unido, tem em seu cerne a operação de aeronaves configuradas em alta densidade de assentos. Busca-se, com isso, acomodar o máximo de passageiros por voo e com isto, minimizar os custos unitários por passageiro. Assim, uma das maiores inovações introduzidas por esse modelo de negócios foi a polêmica redução do pitch dos assentos por meio da adição de fileiras em aeronaves para atingir sua configuração máxima de transporte, dentro dos limites impostos pela regulação. Sinalizando um aprofundamento dessas estratégias, em 2010, a Ryanair conduziu uma pesquisa com 120.000 pessoas, em um estudo que concluiu que um terço delas considerariam adquirir passagens para voos em "assentos verticais" nas aeronaves, caso as passagens fossem grátis, enquanto 42% disseram que os usariam se o preço das passagens fosse a metade de uma passagem com voos em assentos convencionais[4]. Como discutiremos mais adiante, o "assento vertical" é um dos conceitos inovadores de design de cabine em consideração na indústria e que permite incrementar até 20% a mais de capacidade de assentos em um jato da aviação comercial[5].

Por outro lado, a fabricante de assentos Zephyr Aerospace conduziu uma pesquisa ao final dos anos 2010 e apontou que 70% dos viajantes da classe econômica estariam dispostos a trocar algumas das vantagens associadas a taxas extras então oferecidas pelas companhias aéreas, como pacotes de alimentação a bordo e bagagem extra, pela possibilidade de "se deitar e dormir"[6]. Temos que uma vertente importante dos esforços inovativos dos fabricantes e designers de assentos aeronáuticos é oferecer soluções que otimizem o número de passageiros transportados em um voo para atender uma demanda das companhias aéreas low cost e de suas rivais que necessitem de maior competitividade, mas também oferecer possibilidades de maior conforto a bordo para a extração de receitas auxiliares e rentabilidade em geral.

No que se segue, apresentamos alguns dos conceitos inovadores de configuração de assentos em aeronaves de passageiros apresentados pelos desenvolvedores nos últimos anos. Alguns desses conceitos vêm sendo discutidos desde o final dos anos 2000. Sabe-se que muitos deles já foram patenteados[7], apesar de não terem sido certificados pelos órgãos aeronáuticos. Os conceitos inovadores variam conforme a probabilidade de serem aceitos em larga escala pelas companhias aéreas e público em geral, dadas as suas vantagens e desvantagens relativas. Alguns são futuristas ao ponto de não se conseguir antecipar nitidamente o perfil do passageiro disposto a aceitá-lo, e sobre como fariam parte das estratégias comerciais das empresas aéreas. É possível que muitos nunca venham de fato a obter certificação, dado os rigorosos requisitos de segurança de voo a serem cumpridos, por exemplo com questões referentes a procedimentos em meio a uma turbulência ou evacuação da aeronave em caso de emergência. Como ilustração desse rigor na avaliação de novos produtos aeronáuticos, pode-se destacar uma frase de um representante da Airbus em 2014 que sugeriu que "*muitos, senão a maioria dos projetos de patentes [referentes a assentos em aeronaves] nunca serão desenvolvidos*"[8]. Entretanto, faz-se necessário investigar seus efeitos para se antecipar a futuras mudanças de paradigma, episódios esses em geral disruptivos e de difícil projeção com base em experiências passadas.

É necessário destacar a incerteza associada ao impacto desses conceitos nos custos operacionais de uma companhia aérea, incluindo consumo de combustível, manutenção e demais componentes. Por outro lado, todos acarretam mudanças relevantes na densidade de assentos, com efeitos sobre custos unitários, conforto, preços, segmentação e lucratividade das companhias. Com alterações no número de fileiras e assentos, torna-se possível discutir de modo mais embasado alguns de seus possíveis impactos operacionais e comerciais no transporte aéreo. Importante notar, entretanto, que determinadas soluções são mais aplicáveis ao segmento internacional do que no doméstico, dadas as diferenças operacionais, de mercado e de perfil dos passageiros.

---

[4] www.dailymail.co.uk.
[5] www.aviointeriors.it/2018/press/aviointeriors-skyrider-2-0. Vide nossa discussão a seguir sobre o "Skyrider".
[6] www.dailyhive.com/mapped/double-decker-zephyr-seat-airplane.
[7] www.today.com/money/stacked-squatting-squished-7-scary-airline-seat-patents-t52241.
[8] www.newsweek.com/saddle-seat-economy-flights-could-have-you-standing-la-new-york-888943.

O primeiro conceito de design de assentos é o "Morph Seating", da empresa Seymour Powell (Figura 5).

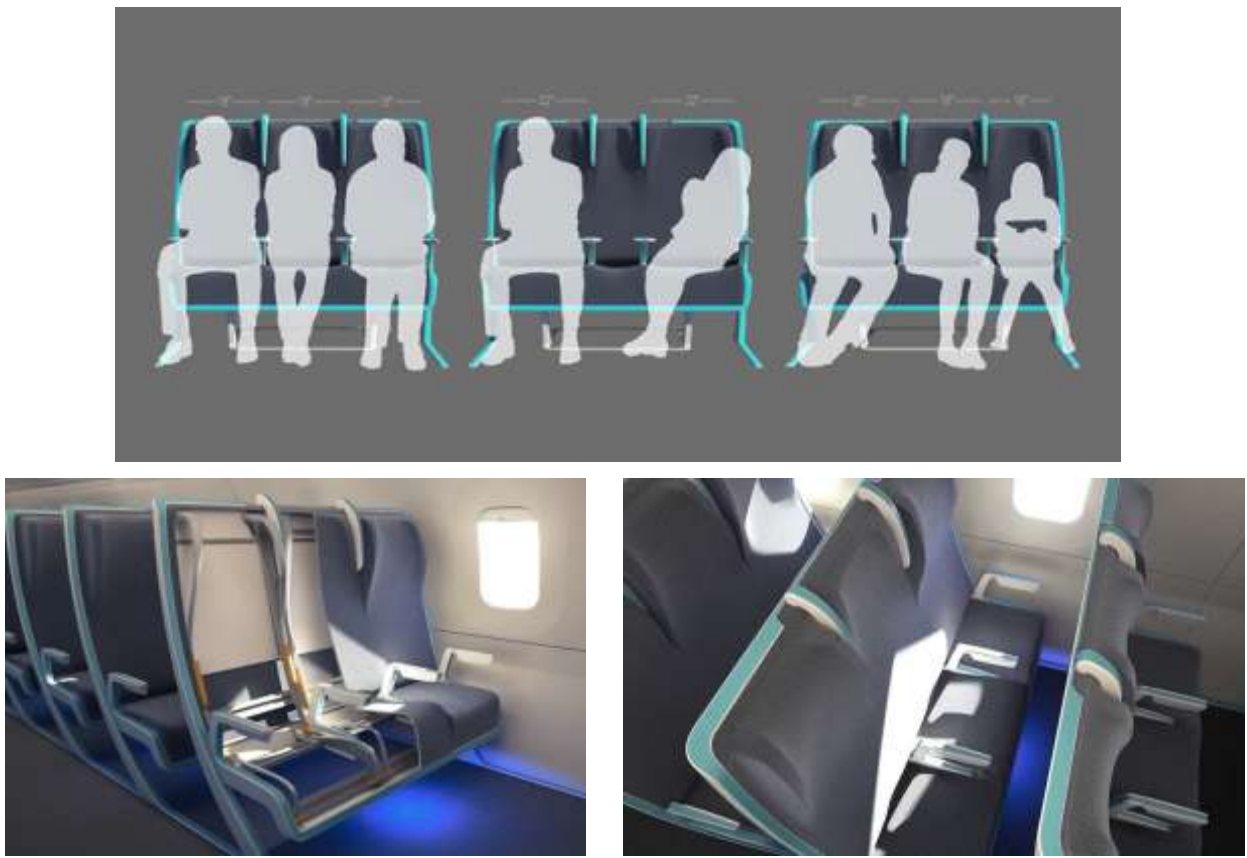

*Fontes: www.seymourpowell.com; www.youtube.com/watch?v=X8REY-oXl3U.*

**Figura 5 - Conceito "Morph Seating" (Seymour Powell)**

O "Morph Seating" possui como ideia principal que, em vez de haver um trio de assentos individuais, existe um tipo de banco fabricado a partir de várias peças de tecido e espuma. Uma única peça de tecido é esticada para formar os assentos e outra compõe o encosto. Os assentos individuais são definidos por apoios de braço e divisórias que fixam o tecido. O conceito baseia-se no uso de tecido esticado e suportes móveis que permitem personalização pelos passageiros e possibilitam às companhias cobrar pela largura adicional do assento entre as amenidades ofertadas no momento da reserva. A flexibilidade da largura é seu principal diferencial.

Do ponto de vista de capacidade e densidade, o "Morph Seating" poderia permitir ajustes na largura conforme a composição de passageiros de cada voo, elevando a densidade quando há maior proporção de viajantes dispostos a aceitar assentos mais estreitos e reduzindo-a quando o foco recai sobre conforto, o que o torna atraente para segmentos com maior disposição a pagar, como viajantes frequentes de negócios, passageiros corporativos e clientes premium em rotas de média duração. Essa flexibilidade permitiria às companhias testar combinações distintas de conforto e capacidade, aplicando tarifas diferenciadas por largura e administrando melhor o equilíbrio entre receita por assento e satisfação do passageiro. Essa mesma flexibilidade, porém, pode introduzir complexidade comercial e operacional, além de risco de frustração caso a percepção de justiça na distribuição de espaço seja afetada, caso a largura contratada não corresponda à experiência real ou caso ocorram conflitos entre passageiros vizinhos com preferências divergentes. A necessidade de ajustes físicos de apoios e divisórias e a exigência de padronização de segurança, limpeza e manutenção podem reduzir parte dos ganhos esperados, de modo que o conceito tende a se adequar mais a nichos ou a cabines híbridas com zonas de maior conforto do que a uma solução aplicável de forma ampla à classe econômica.

Outro conceito introduzido ao longo dos últimos tempos é o da empresa Thompson Aero Seating, a chamada "Cozy Suite" (Figura 6)

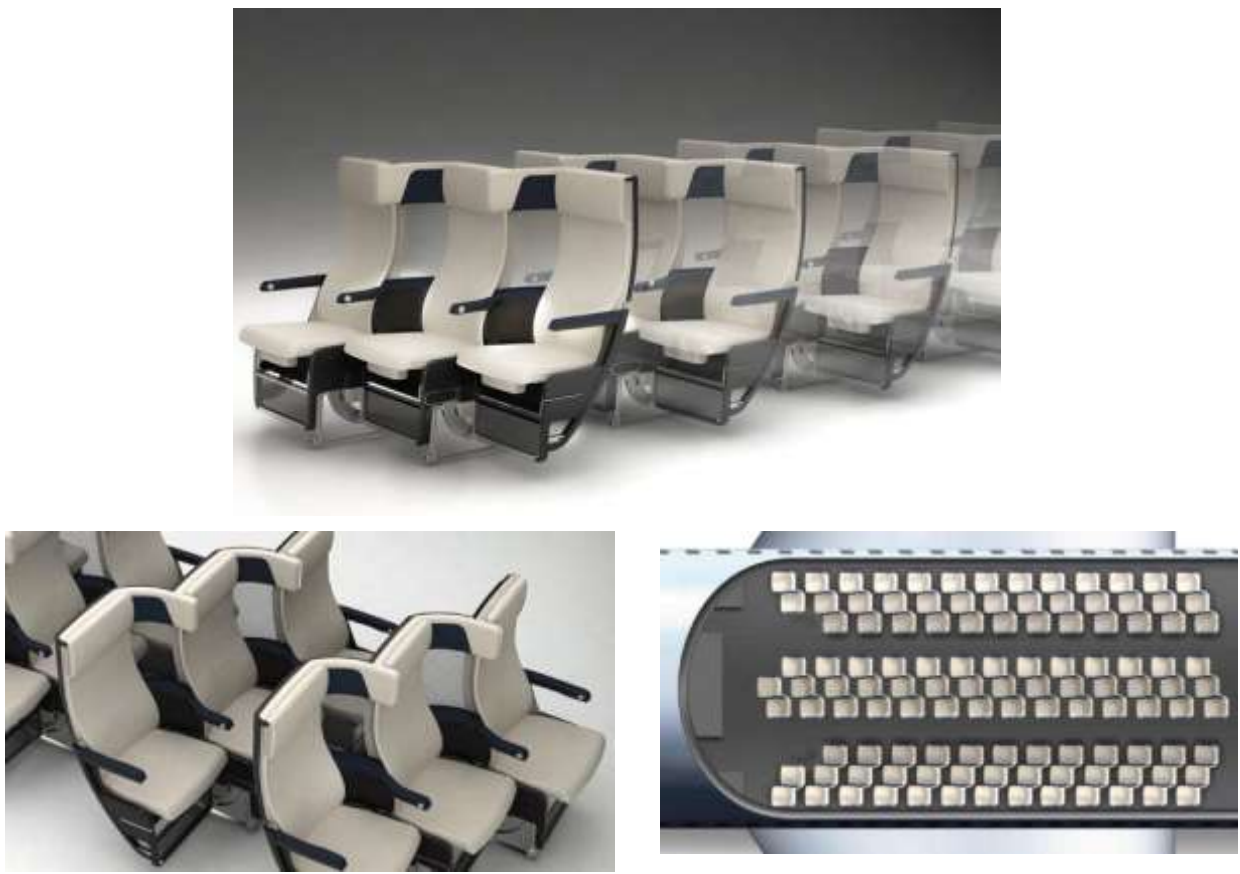

*Fonte: www.thompsonaero.com*

**Figura 6 - Conceito "Cozy Suite" (Thompson Aero Seating)**

Trata-se de um design de assentos que incorpora um segundo apoio de cabeça lateral, além dos apoios de braço individuais. Em tese, ofereceria solução para o conflito gerado pelo uso compartilhado dos apoios de braço, além de proporcionar assento mais largo e acesso facilitado aos corredores. O desenvolvedor indica possibilidade de aumento da densidade de assentos. Em termos de vantagens comparativas, o conceito beneficia de modo evidente o passageiro do assento do meio em relação ao layout atualmente empregado na aviação comercial.

A "Cozy Suite" apresenta implicações diretas para capacidade, densidade e segmentação de mercado. Ao incluir um segundo apoio de cabeça lateral e reorganizar o espaço relativo entre os três passageiros, o conceito pode elevar a densidade sem comprometer a experiência dos ocupantes. Essa reorganização cria espaço para novos níveis de produto dentro da classe econômica, permitindo comercializar zonas com maior privacidade e ergonomia ou aplicar tarifas diferenciadas ao assento do meio redesenhado. Em termos de segmentação, surgem possibilidades de direcionamento a viajantes que valorizam privacidade lateral e passageiros corporativos que preferem um assento que reduza movimentos laterais do corpo ao longo do voo. A promessa de aumento de densidade, porém, exige cautela, pois rearranjos estruturais podem gerar desafios de certificação, ergonomia e evacuação, além de conflitos na percepção de equidade entre passageiros que pagam por atributos de difícil padronização. Existe também o risco de que o ganho de conforto seja percebido como excessivamente assimétrico, favorecendo de modo desproporcional um tipo de assento e criando comparações que podem gerar insatisfação, o que limita a adoção ampla do conceito e aponta para uso em segmentos específicos ou em zonas mais qualificadas dentro da classe econômica.

A Figura 7 apresenta o conceito do "Air Lair". O Air Lair é um espaço na forma de "casulo individual" para o passageiro com uma configuração de dois andares. Trata-se de um assento plano ergonomicamente projetado, que se assemelha a um assento em um automóvel baixo, para fornecer o conforto ao passageiro. O fabricante calcula que a configuração acomoda 30% mais passageiros dentro do mesmo espaço na cabine[9]. Dentro do ambiente isolado do casulo, o passageiro pode fazer ajustes para controlar seu próprio espaço pessoal sem afetar outros passageiros. O fabricante propõe o uso de uma iluminação estrategicamente posicionada para criar um ambiente que estimule uso de equipamentos de entretenimento a bordo e, assim, geração adicional de receitas auxiliares.

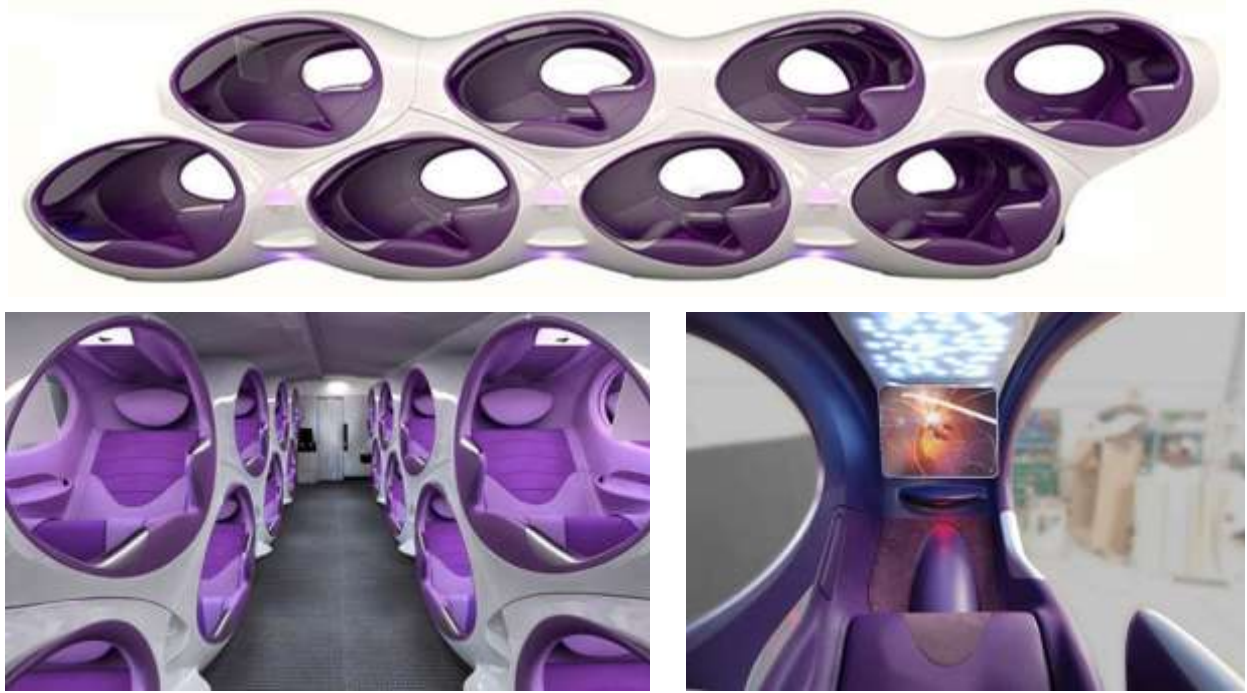

*Fontes: www.collabcubed.com/2012/06/13/air-lair-pod-business-class-seats;*
*www.dailymail.co.uk/travel/travel_news/article-2839713/Take-skies-double-decker-sleep-pods-Futuristic-stacked-cocoons-FINALLY-air-passengers-privacy.html.*

**Figura 7 - Conceito "Air Lair" (Factorydesign)**

O "Air Lair" altera de modo mais radical a relação entre capacidade, densidade e segmentação ao transformar o espaço em casulos empilhados em dois níveis e ao oferecer isolamento individual que reduz conflitos por espaço e interação entre passageiros. A possibilidade de acomodar cerca de 30 por cento mais passageiros cria uma vantagem de densidade relevante, sobretudo em rotas internacionais de longa duração, onde o custo do espaço por passageiro é elevado. Para a companhia aérea, o modelo permite segmentação aprofundada, com casulos diferenciados por nível, posição, privacidade, iluminação e integração com entretenimento, aproximando a lógica da cabine de produtos modulados presentes em serviços premium. O formato de casulo, porém, pode gerar barreiras de aceitação entre passageiros que rejeitam ambientes mais fechados, além de desafios regulatórios relacionados a evacuação, circulação e acessibilidade. Existe também o risco de percepção de desigualdade caso diferenças entre casulos sejam difíceis de justificar por meio de tarifas ou caso o aumento de densidade comprometa a sensação geral de espaço. Assim, o conceito tende a se alinhar a nichos específicos ou a cabines mistas voltadas à maximização de receita em voos internacionais, em vez de se configurar como solução aplicável de forma ampla à classe econômica.

---

[9] www.factorydesign.co.uk/aviation/case-study-aviation-air-lair,

A Figura 8 apresenta imagens do "StepSeat", da empresa Jacob-Innovations. O StepSeat é composto de assentos elevados alternados, de forma a aproveitar espaço vertical no interior de aeronaves por meio do uso de degraus. A ideia do conceito é proporcionar maior conforto ao passageiro, mantendo a mesma densidade de assentos a bordo[10].

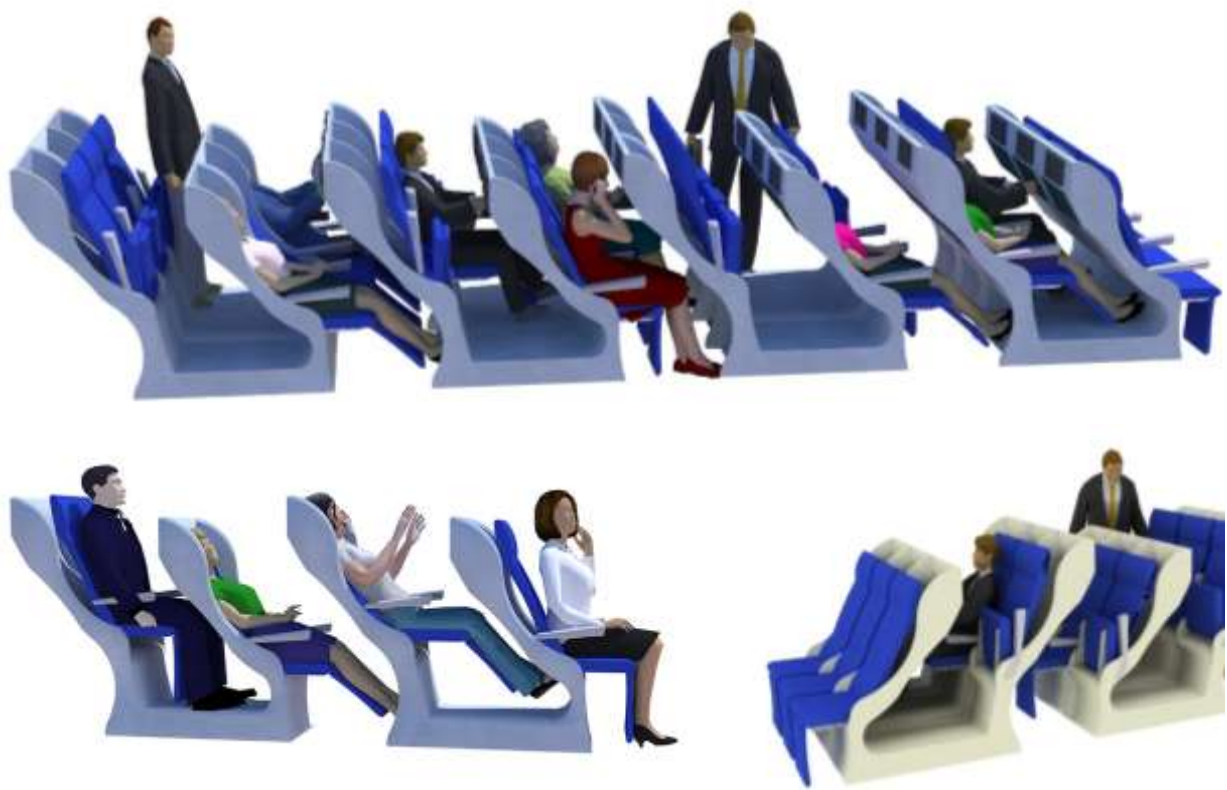

*Fonte: www.jacob-innovations.com.*

**Figura 8 - Conceito "StepSeat" (Jacob-Innovations)**

O "StepSeat" altera a organização espacial ao introduzir assentos elevados de forma alternada, utilizando o volume vertical sem modificar a quantidade total de lugares. Em princípio, isso permitiria elevar o conforto relativo por passageiro sem sacrificar densidade, criando um produto que poderia ocupar posição intermediária entre o assento padrão e alternativas mais caras de conforto. Para a segmentação de mercado, a companhia aérea poderia explorar categorias diferenciadas segundo altura, privacidade e ergonomia, direcionando o produto a passageiros que valorizam sensação de espaço pessoal, viajantes corporativos em rotas de média duração ou clientes dispostos a pagar por uma experiência menos convencional. Uma análise crítica, porém, indica desafios como possíveis dificuldades de evacuação, segurança no uso dos degraus, padronização do atendimento e aceitação do público, especialmente entre passageiros com mobilidade reduzida ou aversão a estruturas elevadas. Existe ainda o risco de percepção muito heterogênea de conforto, o que torna o produto mais adequado a nichos ou a seções restritas da cabine, sobretudo em operações internacionais, onde há maior tolerância a inovações estruturais e maior disposição a pagar por diferenciação.

---

[10] www.jacob-innovations.com/Economy-Comfort.html.

Outro conceito da Jacob-Innovations que utiliza degraus para capturar o espaço vertical não utilizado nos assentos atuais é o FlexSeat. (Figura 9). O FlexSeat utiliza a ideia de avião de "dois andares", conceito já utilizado nos maiores aviões Boeing 747 e Airbus A380, além de alguns modelos de ônibus e trens. O fabricante enfatiza não apenas a característica de assentos com reclinação superior aos atuais, mas recursos superiores mesmo à Classe Executiva convencional, como privacidade, a possibilidade de viajar com bebês, e espaço para bagagem de mão de maior porte. Adicionalmente, o conceito permite uma conversão "instantânea" da Classe Executiva para a Classe Econômica[11], por meio de ajustes nos degraus para o nível superior. Essa característica permitiria às companhias aéreas vender bilhetes para as duas classes de acordo com a demanda.

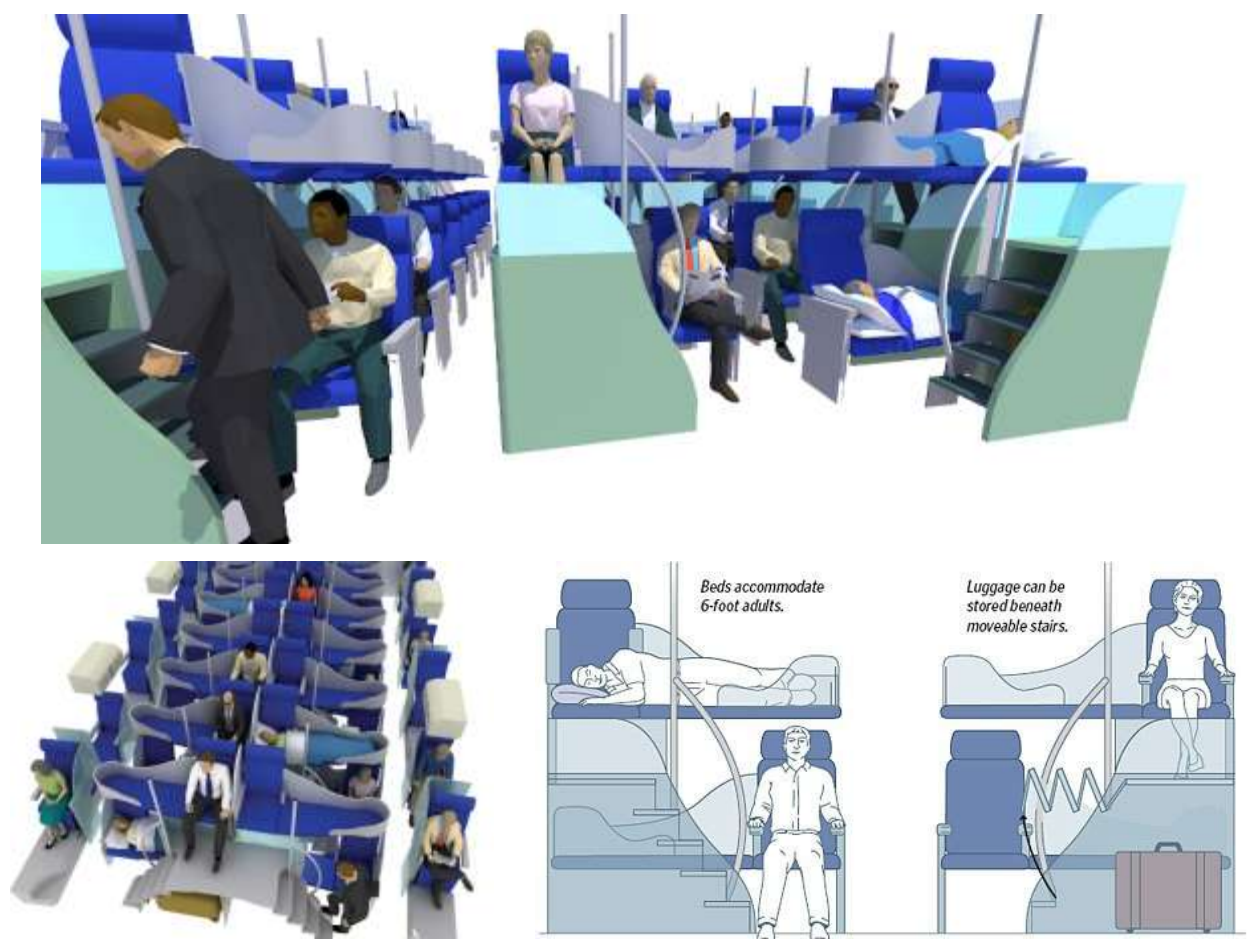

*Fonte: www.jacob-innovations.com.*

**Figura 9 - Conceito "FlexSeat" (Jacob-Innovations)**

O "FlexSeat" amplia o uso do espaço vertical ao adotar uma lógica de dois andares inspirada em aeronaves de grande porte, ônibus e trens, oferecendo reclinação superior, maior privacidade, possibilidade de acomodar bebês e espaço adicional para bagagem de mão. A conversão rápida entre Classe Executiva e Econômica cria um instrumento de gestão de capacidade que permite ajustar a oferta conforme a demanda, ampliando a flexibilidade comercial e o leque de produtos tarifários. Do ponto de vista de segmentação, a companhia aérea poderia explorar diferentes níveis dentro de cada classe, distinguindo posições superiores e inferiores, graus de privacidade e layouts voltados a famílias, viajantes corporativos ou passageiros premium em rotas internacionais. Uma análise mais cuidadosa, porém, aponta desafios estruturais e regulatórios, como certificação de segurança, evacuação, acessibilidade e manutenção de uma cabine de dois níveis em espaço reduzido. A heterogeneidade de preferências pode limitar a aceitação do produto, e a complexidade operacional relacionada à conversão entre classes pode reduzir parte dos ganhos previstos, o que sugere que

---

[11] www.jacob-innovations.com/FlexSeat.html.

o conceito tende a se adequar a nichos ou a operações internacionais com maior tolerância a soluções inovadoras e maior disposição a pagar.

Uma outra proposta de design de assentos de avião de dois andares é da empresa Zephyr Aerospace - o "Zephyr Seat" ou "Assento plano acessível para avião" (do inglês "Affordable lie-flat airplane seating"), apresentado na Figura 10. O desenvolvedor aponta que o conceito pode ser adaptado em áreas já existentes da Classe Econômica ou Econômica Premium das aeronaves sem perder os assentos atuais, ampliando a capacidade de 20% em aeronaves de longo curso.

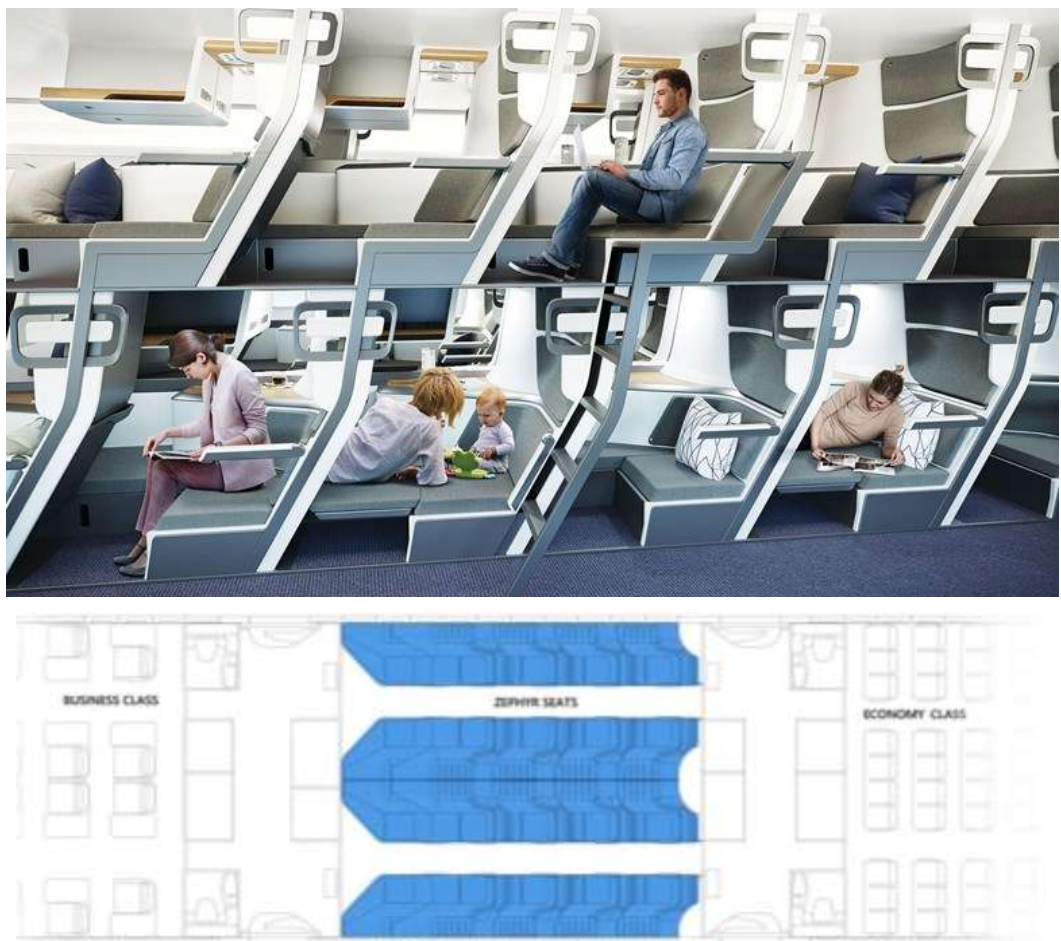

*Fontes: www.linkedin.com/company/zephyrseat/about/; www.dailyhive.com/mapped/double-decker-zephyr-seat-airplane; www.republic.co/zephyr-aerospace.*

**Figura 10 - Conceito "Zephyr Seat", "Affordable lie-flat airplane seating" (Zephyr Aerospace)**

O "Zephyr Seat" propõe uma configuração de dois andares adaptável às áreas existentes da Classe Econômica ou Econômica Premium, mantendo os assentos atuais e acrescentando lugares adicionais no nível superior, o que pode elevar a capacidade em cerca de 20 por cento em aeronaves de longo curso. Para a companhia aérea, essa expansão abre espaço para segmentação mais precisa, permitindo ofertar posições superiores como produto intermediário entre Econômica Premium e Executiva, com tarifas definidas segundo privacidade, reclinação e conforto térmico. A estrutura possibilita ainda vendas direcionadas a perfis específicos, como viajantes corporativos sensíveis a preço, passageiros de longa duração que buscam descanso horizontal ou clientes que valorizam isolamento relativo em cabines densas. Uma análise crítica, porém, aponta desafios como requisitos de certificação, evacuação, estabilidade estrutural e aceitação do público, especialmente entre passageiros que associam o nível superior a confinamento ou maior dificuldade de acesso.

A Figura 11 apresenta o conceito da Zodiac Seats, de "Cabin Hexagon". A ideia dos hexágonos consiste em assentos alternados voltados para a frente e para trás, de forma a melhorar o uso do espaço da cabine. Essa disposição não convencional dos assentos visa otimizar o número de passageiros transportados em um voo, evitando o contato lateral entre passageiros[12]. Por outro lado, possui a característica que pode ser indesejável de reduzir a privacidade pelo maior contato frontal.

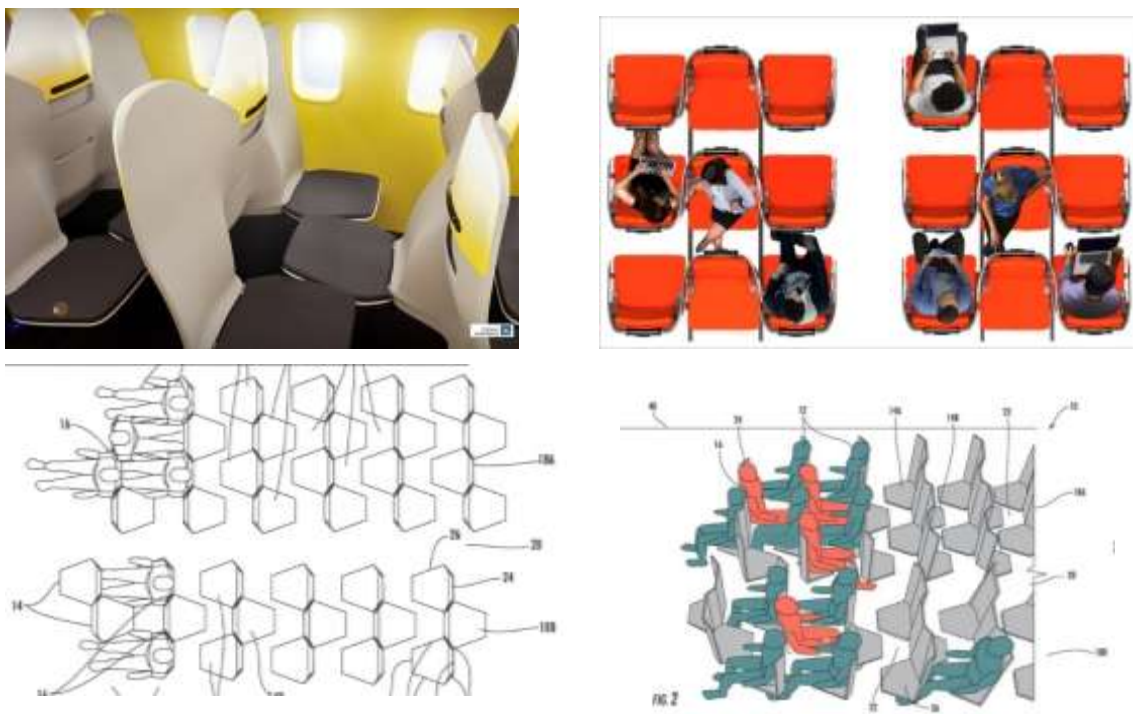

*Fontes: www.flyertalk.com/articles/economy-class-cabin-hexagon-promises-more-seats-more-legroom-but-at-will-it-fly.html; www.bluebus.com.br/esse-layout-de-assentos-promete-tornar-as-viagens-de-aviao-ainda-mais-dificeis; www.fortune.com/2015/07/09/new-airplane-seating-hexagon;www.paxex.aero/aviointeriors-janus-economy-class-yin-yang.*

**Figura 11 - Conceito "Cabin Hexagon" (Zodiac Seats)**

Assim, o "Cabin Hexagon" reorganiza o espaço ao alternar assentos voltados para a frente e para trás, buscando otimizar a ocupação da cabine e eliminar o contato lateral entre passageiros. Essa geometria permite elevar a eficiência espacial e, em alguns arranjos, aumentar a capacidade sem comprometer o conforto individual, criando oportunidades de segmentação por posição, já que determinados ângulos oferecem sensação ampliada de espaço ou melhor acesso visual ao corredor. A companhia aérea poderia explorar essas diferenças tarifando posições mais demandadas, oferecendo categorias intermediárias entre o assento padrão e versões de maior conforto ou direcionando o produto a passageiros menos sensíveis à privacidade lateral. Uma análise crítica, porém, indica limitações importantes. A proximidade frontal pode reduzir a privacidade, o que pode dificultar a aceitação, especialmente entre passageiros que evitam contato visual prolongado. Existem também questões de certificação, ergonomia e gestão do fluxo de cabine, além da possibilidade de percepção desigual de conforto entre as posições do hexágono.

A Figura 12 apresenta o conceito de "Checkerboard" da empresa Butterfly. Nele, propõe-se a configuração de assentos alternados em ambos os eixos x e y da cabine, em um padrão xadrez. Os assentos podem ser dobrados para dar aos assentos adjacentes maior espaço para os cotovelos - até duas polegadas a mais para assentos do corredor, e quatro polegadas extras para os assentos do meio -, além de 20 centímetros (quase 8 polegadas) de espaço extra para as pernas para os assentos imediatamente atrás[13]. Esse "tabuleiro de damas" é um sistema de assentos flexível que permite fácil conversão entre a Classe Econômica de alta densidade e a Classe Executiva, em voos de curta e média distâncias. O fabricante enfatiza a característica de "conversão instantânea", que pode ser feita rapidamente pela tripulação de cabine antes de um voo.

---

[12] www.dailymail.co.uk/travel/travel_news/article-3154771/Thought-economy-class-couldn-t-worse-horrifying-hexagonal-seating-means-FACE-passengers.html.
[13] www.butterflyseating.com/checkerboard.

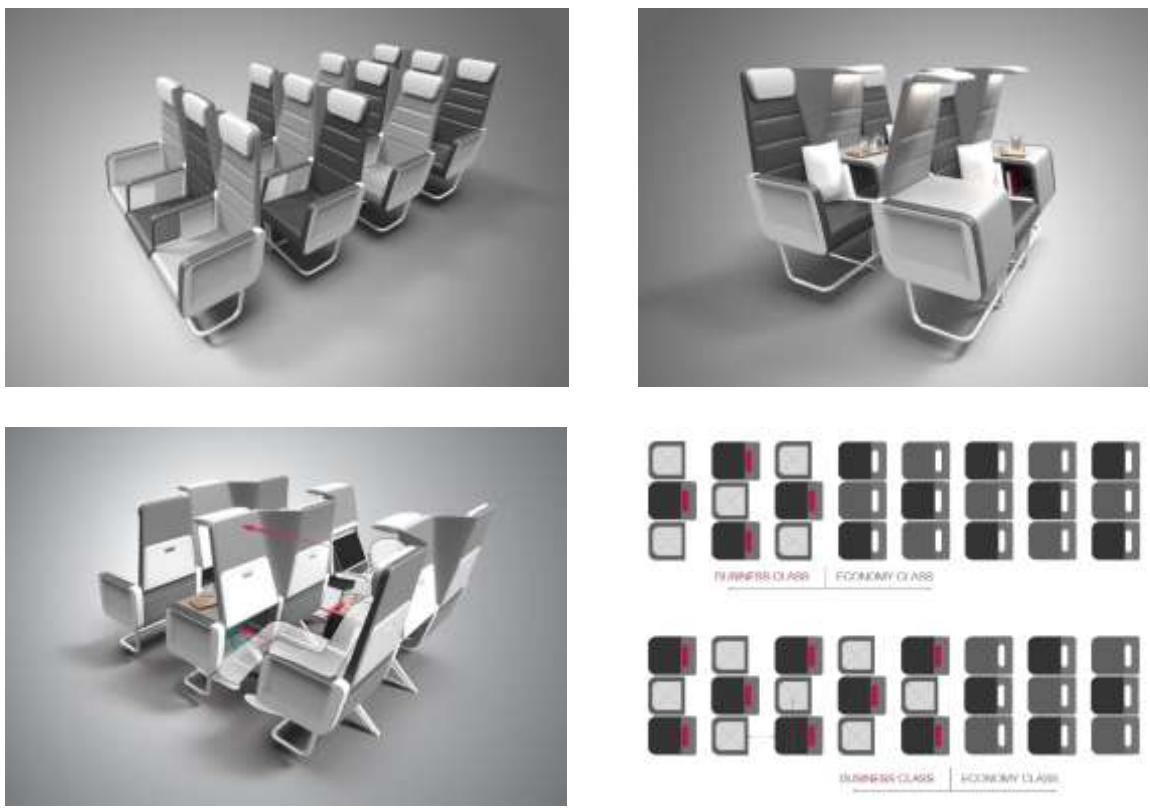

*Fonte: www.butterflyseating.com/checkerboard.*

**Figura 12 - Conceito "Checkerboard Convertible Seating System" (Butterfly)**

Pode-se perceber que a proposta do "Checkerboard" reorganiza a cabine em padrão xadrez nos eixos x e y, permitindo que assentos adjacentes sejam dobrados para ampliar espaço para cotovelos e pernas, criando uma geometria que redistribui o conforto sem reduzir o número total de lugares. Em termos de segmentação, essa flexibilidade abre espaço para produtos intermediários dentro da Classe Econômica, com tarifas diferenciadas segundo posição, amplitude de movimento, privacidade lateral e espaço para as pernas, além de possibilitar versões compactas da Classe Executiva em rotas de curta e média distâncias. A conversão rápida entre alta densidade e uma configuração mais confortável oferece às companhias um instrumento para ajustar a oferta conforme a demanda, especialmente em mercados internacionais com maior heterogeneidade de perfis e maior disposição a pagar. Uma análise crítica, porém, indica limitações como risco de percepção de desigualdade entre posições, dificuldade de gestão de expectativas quando o espaço adicional depende de assentos dobrados e desafios de certificação, manutenção e padronização operacional. Existe ainda o risco de que a experiência seja percebida como inconsistente se a configuração variar de modo frequente entre voos, o que sugere adequação maior a nichos ou a operações internacionais que toleram melhor soluções estruturais diferenciadas.

Completamos a exposição dos modelos invasores com o mais polêmico deles, o Skyrider, da Aviointeriors (Figura 13). O Skyrider, que está em sua versão 3.0, é um assento que inova ao radicalizar no uso do espaço interno da cabine, permitindo um tipo de densidade "ultraelevada" (ultra-high seat density). Sua principal característica é o formato original tipo selim de bicicleta[14], com o objetivo de manter a postura ereta do passageiro em uma configuração com inclinação e pitch reduzidos de 23 polegadas. Segundo o fabricante, o design deste assento permite aumentar o número de passageiros a bordo em 20%, além de pesar 50% menos do que os assentos convencionais da classe econômica padrão[15]. O Skyrider é polêmico por reforçar ao extremo a ideia de perda de espaço para as pernas dos passageiros, o que é visto com desconfiança na indústria e no público em geral. Mas não pode deixar de ser qualificado como inovador, ao apresentar uma nova fronteira de passagens de baixo custo e potencialmente abrir a possibilidade de voar a outros tipos de passageiros, alguns dos quais atualmente não podem pagar pelas passagens aéreas, mas que poderiam ingressar no mercado caso os preços fossem realmente muito baixos.

---

[14] Há notícia que em 2014 a Airbus também tenha patenteado o seu próprio design de assento em formato de selim bicicleta (fonte: www.newsweek.com/saddle-seat-economy-flights-could-have-you-standing-la-new-york-888943). Importante salientar que o setor de manufatura de aeronaves é particularmente suscetível à circulação de informações não verificadas e que mesmo veículos de comunicação consolidados podem, em alguns casos, reproduzir notícias imprecisas ou não confirmadas, já que projetos preliminares, estudos conceituais e protótipos exibidos em feiras podem gerar interpretações exageradas sobre sua adoção futura.

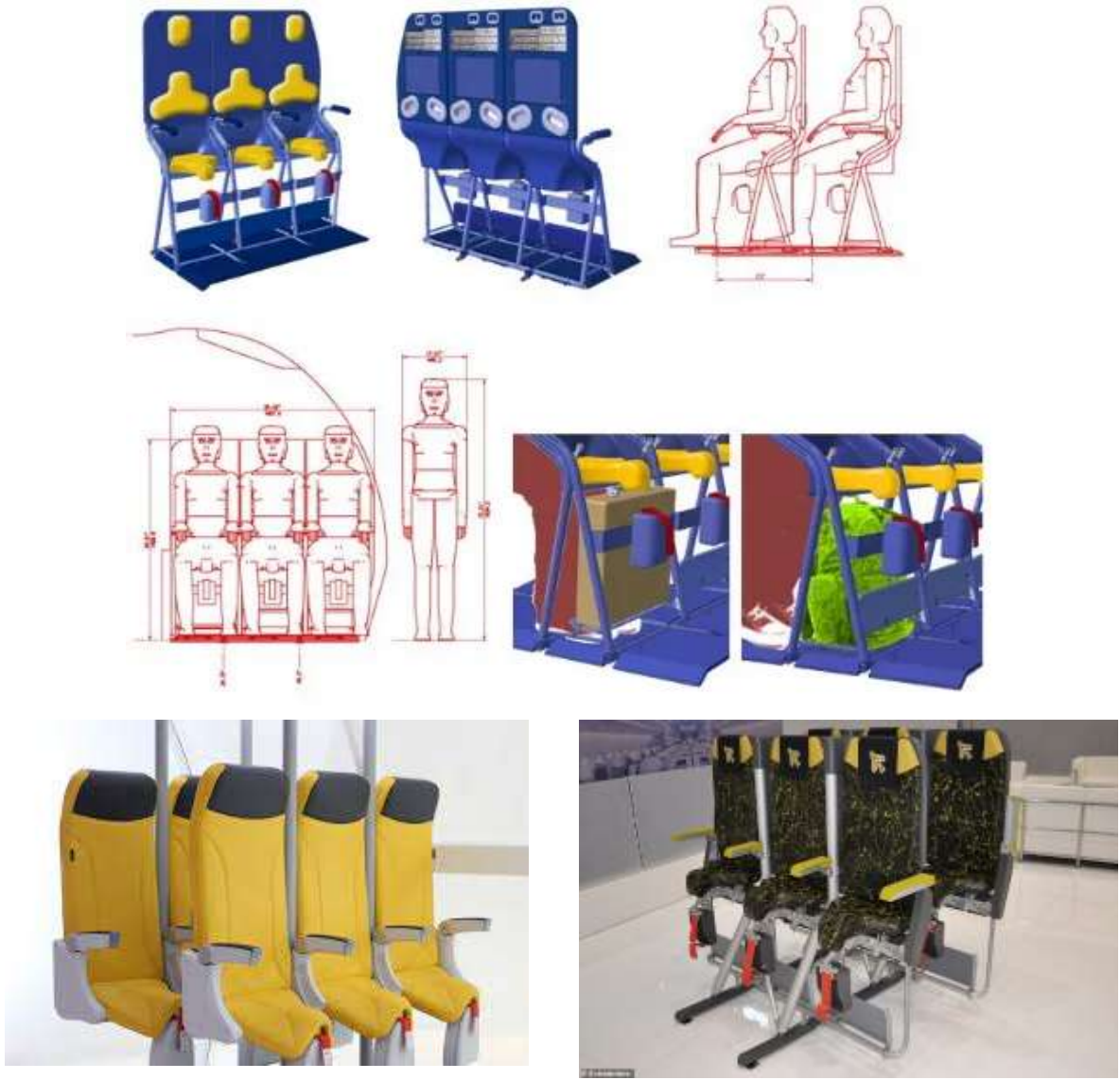

*Fontes: www.aviointeriors.it/2018/press/aviointeriors-skyrider-2-0; www.dailymail.co.uk/travel/travel_news/article-6885907/Standing-planes-moves-step-closer-reality-thanks-Aviointeriors-Skyrider-seat.html;*

**Figura 13 - Conceito "Skyrider" (Aviointeriors)**

O histórico do projeto vem de 2010, quando a versão inicial foi apresentada e divulgada de forma a apresentar o formato tipo selim como uma solução para reduzir o espaço necessário para cada ocupante e, em tese, permitir ampliar a capacidade de aeronaves de corredor único, especialmente em rotas curtas. A partir de 2018 surgiram versões atualizadas identificadas como 2.0, com reforço estrutural, maior estabilidade do encosto e pitch de aproximadamente 23 polegadas. Versões mais recentes, como a 3.0, incorporaram simplificações, redução de peso e pequenas adaptações ergonômicas, mantendo a premissa de elevar a capacidade em cerca de 20 por cento e reduzir o peso em torno de metade em relação aos assentos tradicionais[16].

---

[15] www.aviointeriors.it/2018/press/aviointeriors-skyrider-2-0.

A discussão sobre o Skyrider ganhou visibilidade quando a Ryanair manifestou interesse em explorar configurações de ultrabaixo custo, mencionando a possibilidade de assentos verticais ou quase em pé. No entanto, não há registro de que qualquer teste operacional tenha sido autorizado ou conduzido. A Ryanair posteriormente negou planos concretos de adoção, embora tenha utilizado o conceito no discurso institucional sobre tarifas muito baixas e como estratégia de atrair engajamento.[17]

A viabilidade do Skyrider enfrenta obstáculos significativos. No plano regulatório, configurações quase verticais levantam questões essenciais de certificação, como evacuação rápida, proteção em turbulências, adequação de cintos de segurança e atendimento a requisitos mínimos de ergonomia.[18] No plano operacional, sua adoção exigiria revisão de procedimentos de embarque, adaptações na lógica de limpeza e manutenção e treinamento específico da tripulação para lidar com um produto distante do padrão atual. No mercado, a aceitação pelos passageiros é incerta, pois o desconforto percebido tende a ser elevado mesmo em voos curtos, e há risco de prejuízo à imagem da companhia que o adotar. A ideia de atrair público disposto a sacrificar conforto extremo em troca de tarifas muito baixas pode ser válida em alguns mercados, mas não resolve a heterogeneidade de preferências e pode gerar efeitos negativos de reputação.

Como resultado, o Skyrider permanece como conceito inovador não testado, cuja adoção ampla encontra barreiras regulatórias, operacionais e comerciais. A proposta ilustra limites extremos de densidade de cabine e de estratégias de ultrabaixo custo, mas sua implementação real, especialmente em aeronaves como os Boeing 737-800 operados pela Ryanair, parece muito improvável no curto e médio prazos.

A consideração conjunta dos conceitos inovadores acima ilustrados, bem como dos resultados da literatura da área (Seção II), permite construir um quadro comparativo mais amplo sobre como densidade de assentos, conforto e custos operacionais interagem no desenho de produtos na aviação comercial. Configurações que elevam a densidade, como Skyrider, Cabin Hexagon ou propostas de dois andares econômicos, seguem lógica de redução de custos unitários por meio da ampliação da capacidade. Já soluções como Morph Seating, Cozy Suite ou Zephyr Seat atuam no sentido oposto, aumentando conforto e criando microprodutos capazes de sustentar diferenciação tarifária em mercados competitivos, em linha com evidências de que melhorias segmentadas podem gerar prêmios ou fortalecer receitas quando o público é menos sensível a preço. Em ambos os casos, mantém-se a ligação entre estrutura de cabine, elasticidades diferenciadas e estratégias de precificação. Assim, tanto os conceitos que elevam a densidade quanto aqueles que ampliam o conforto mostram que o layout interno é instrumento econômico que molda custos operacionais, qualidade percebida e formas de segmentação tarifária.

O Skyrider tende a ser menos viável que os demais conceitos analisados. Embora vários projetos enfrentem desafios de certificação, aceitação e complexidade operacional, o Skyrider posiciona esse conjunto de dificuldades em nível mais crítico, apesar de ser provavelmente o que mais atrai a atenção pública. A razão da sua inviabilidade é que o conceito modifica não apenas a geometria da cabine, mas a própria definição de assento, aproximando a experiência de postura quase vertical com apoio tipo selim. Como discutido acima, isso contraria normas de segurança, ergonomia mínima, requisitos de evacuação e proteção contra impactos, criando barreiras mais rígidas do que as encontradas em conceitos de dois andares, casulos ou arranjos alternados de poltronas. Além disso, outros projetos oferecem algum ganho perceptível de conforto, privacidade ou modularidade, mesmo quando associados a maior densidade. O Skyrider, ao contrário, enfatiza o aumento extremo de capacidade com perda explícita de conforto, o que dificulta sua compatibilidade com a percepção de valor do passageiro e com a imagem da companhia. Essa assimetria limita a aceitação, reduz a disposição a pagar e restringe o público-alvo a passageiros extremamente sensíveis a preço em rotas curtas.

A Tabela 1 apresenta uma síntese comparativa dos conceitos inovadores de layout de cabine discutidos nesta seção, organizando informações sobre fabricante, data aproximada de apresentação ao mercado, segmento mais adequado e potenciais efeitos sobre densidade de assentos e segmentação comercial.

---

[16] https://www.businessinsider.com/skyrider-standing-airplane-seats-claims-makes-flights-cheaper-2018-4 e diversas outras notícias.

[17] https://www.euronews.com/business/2025/05/23/ryanair-denies-claims-flights-will-soon-offer-cheaper-standing-seats.

[18] https://www.airlineratings.com/articles/we-call-bs-on-these-airline-seats.

**Tabela 1 - Comparação dos principais conceitos inovadores de layout de cabine segundo densidade e segmentação**

| Conceito | | Fabricante | Lançamento aproximado | Segmento mais apropriado | Potencial efeito na densidade | Potencial efeito na segmentação |
|---|---|---|---|---|---|---|
| Morph Seating | 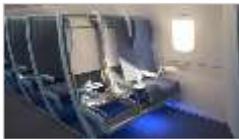 | Seymourpowell | 2013 | Doméstico ou Internacional | Médio | Médio |
| Cozy Suite | 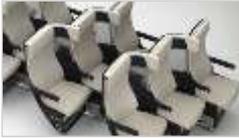 | Thompson Aero Seating | 2008 | Doméstico ou Internacional | Médio | Médio |
| Air Lair | 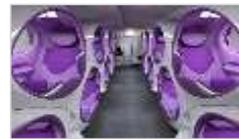 | Factorydesign | 2012 | Internacional (longo curso) | Alto | Alto |
| StepSeat | 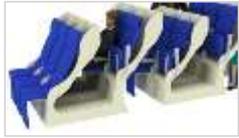 | Jacob-Innovations | 2014 | Doméstico ou Internacional (rotas curtas e médias) | Baixo | Alto |
| FlexSeat | 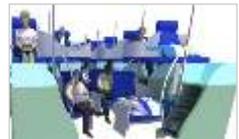 | Jacob-Innovations | 2006 | Internacional (longo curso) | Alto | Alto |
| Zephyr Seat | 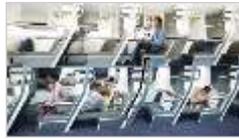 | Zephyr Aerospace | 2020 | Internacional (longo curso) | Alto | Alto |
| Cabin Hexagon | 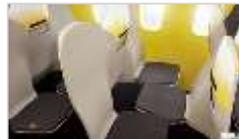 | Zodiac Seats (Zodiac Aerospace France) | 2015 | Doméstico ou Internacional | Médio | Médio |
| Checkerboard | 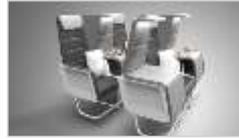 | Butterfly / Paperclip Design | 2012 | Doméstico ou Internacional | Baixo | Médio |
| Skyrider | 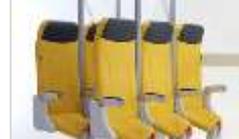 | Aviointeriors | 2010 | Doméstico (rotas curtas) | Alto | Baixo |

*Fontes: Diversos websites especializados em aviação comercial, páginas institucionais dos fabricantes, reportagens online do setor, Wikipedia. Classificações qualitativas próprias dos autores.*

Na Tabela 1, as classificações dos potenciais efeitos na densidade e na segmentação (últimas colunas da direita) são dos autores, baseadas na análise das características estruturais de cada conceito e nos custos operacionais associados, tal como discutido em diversas fontes técnicas e setoriais. Os conceitos classificados como de impacto alto sobre densidade, como Air Lair, Zephyr Seat e Skyrider, compartilham o uso intensivo do espaço vertical ou a redução do espaço individual, mesmo com estratégias radicalmente opostas, mas permitindo acrescentar novos níveis ou compactar o assento a ponto de criar capacidade adicional significativa. Já os conceitos de impacto médio, como Morph Seating, Cozy Suite, FlexSeat e Cabin Hexagon, promovem rearranjos estruturais que melhoram a eficiência espacial ou introduzem novas geometrias, mas sem alterar de modo tão profundo a arquitetura interna da cabine. Por fim, StepSeat e Checkerboard permanecem com impacto baixo, pois preservam o número total de assentos, atuando principalmente na redistribuição de conforto, ergonomia ou privacidade lateral. A classificação comparativa permite identificar quais propostas se orientam à maximização de capacidade e quais privilegiam diferenciação de produto voltada à segmentação tarifária.

### III.3. Mapeamento Exploratório de Configurações Alternativas de Cabine

A partir do conjunto de conceitos inovadores de assentos apresentados anteriormente, esta seção tem por objetivo ilustrar, por meio de um exemplo, algumas possibilidades de arranjo interno de cabine no contexto da aviação comercial doméstica brasileira. Busca-se mostrar, de forma visual e direta, como alterações no pitch, na distribuição de classes econômicas e no número de fileiras podem modificar a densidade e a organização espacial da cabine. Trata-se de um exercício descritivo, que ajuda a compreender como variações estruturais no layout podem dialogar, em etapas posteriores, com temas como conforto, diferenciação de produto e estrutura tarifária.

A Figura 14 apresenta três configurações alternativas de um Boeing 737-800 utilizado em operações domésticas. Na figura, a notação (P, R, S) indica, para uma determinada área da cabine, respectivamente, o pitch do assento em polegadas P, o número de fileiras de assentos R e o total de assentos S. Assim, por exemplo, (P, R, S) igual a (31, 31, 183) deve ser lido como pitch de 31 polegadas, 31 fileiras e 183 assentos disponíveis. As duas primeiras configurações, mostradas na parte superior da figura, foram efetivamente utilizadas pela Gol em 2014 e apresentam capacidades para 183 e 177 passageiros. Em ambos os casos, o pitch principal é de 31 polegadas. Na configuração com 177 assentos existem duas classes econômicas, denominadas "CE/C" (Classe Conforto-Econômica) e "SE/C" (Classe Econômica Padrão, ou "Standard"). Na CE/C o pitch é de 34 polegadas, abrangendo 42 assentos, isto é, (P, R, S) igual a (34, 7, 42).

A terceira configuração, apresentada na parte inferior da Figura 14, é fictícia e serve apenas para fixar ideias sobre o papel da densidade de fileiras no interior da cabine. Trata-se de uma representação ilustrativa, inteiramente desvinculada de qualquer análise real de peso, balanceamento, certificação ou engenharia de manufatura de aeronaves, razão pela qual sugerimos fortemente sua utilização apenas como referência visual, sendo necessários estudos futuros de viabilidade técnica. Nessa configuração, mantém-se a CE/C com pitch de 34 polegadas, preserva-se a SE/C com pitch de 31 polegadas e insere-se uma terceira classe econômica, "BE/C" (Classe Econômica Básica), baseada no conceito do assento Skyrider (vertical, em forma de selim), representada em roxo. Nessa classe, o pitch seria de 23 polegadas, conforme indicado pelo desenvolvedor. A inclusão dessa classe elevaria a capacidade total para 195 passageiros. Mesmo havendo discutido os desafios e mesmo viabilidade desse tipo de configuração de fileiras de poltronas na aviação comercial moderna, a Figura 14 permite visualizar de modo imediato como diferentes propostas de layout alteram densidade, ocupação e distribuição interna, auxiliando na compreensão do problema e na interpretação das relações entre capacidade, conforto e organização espacial da cabine.

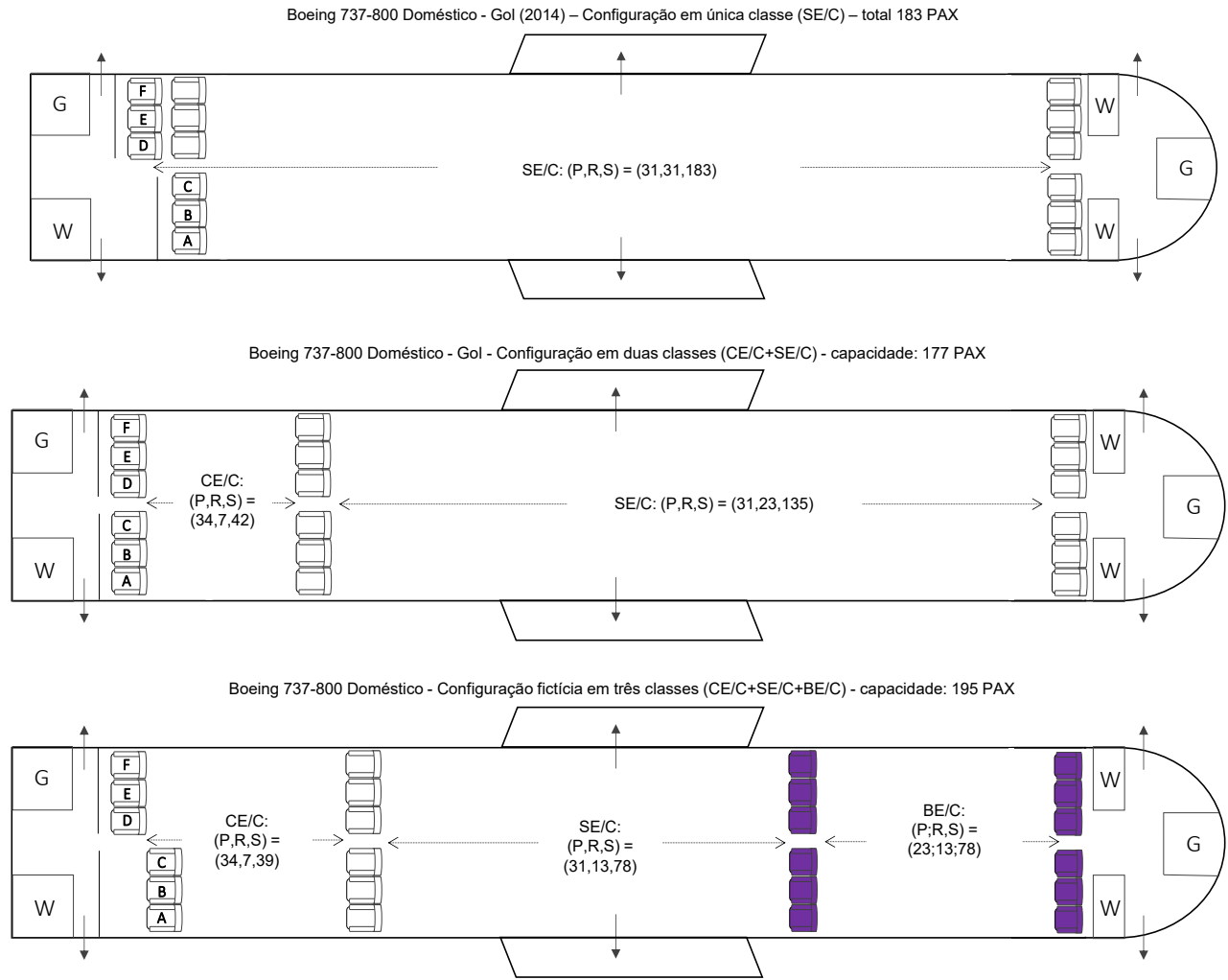

*Nota: SE/C = classe "Standard Economy"; CE/C = classe "Confort Economy"; BE/C = Classe "Basic Economy"; P, R e S denotam, respectivamente, o pitch do assento, o número de fileiras de assentos e o total de assentos na cabine.*

**Figura 14 - Configurações de aeronaves - Gol e exemplo fictício de introdução de conceito inovador de layout**

# IV. Modelo Conceitual

O estudo dos efeitos de mercado das decisões de layout de cabine exige uma compreensão ampla do ambiente competitivo do transporte aéreo. Essas decisões envolvem, de um lado, o adensamento das cabines, isto é, quantas fileiras e qual nível de espaço entre assentos serão oferecidos, e, de outro, a segmentação de passageiros, definida pela criação ou não de subprodutos dentro da classe econômica, como assentos premium ou áreas diferenciadas na aeronave. Essas escolhas se articulam com a demanda por viagens, com as estratégias de precificação das passagens aéreas, com a geração de receitas auxiliares e com a estrutura de custos operacionais das empresas. Em conjunto, esses elementos moldam o posicionamento competitivo de cada companhia aérea e influenciam diretamente o perfil de passageiros que ela é capaz de atrair e reter. Diante da complexidade do problema, a presente seção visa propor um arcabouço conceitual que sirva de base e orientação para as análises empíricas a serem desenvolvidas.

A Figura 15 apresenta um diagrama em forma de árvore de decisão, estruturado como um jogo sequencial, que permite visualizar de maneira o processo decisório das companhias aéreas no que diz respeito à configuração interna das cabines de suas aeronaves, bem como as implicações que essas decisões possuem para a configuração de seu modelo de negócios. Denominamos o diagrama de "Jogo decisório de configuração de cabine de aeronaves e posicionamento de mercado de companhias aéreas", que serve de ilustração da interação estratégica envolvida e que norteia nossas análises. Note que esse ambiente decisório está envolto na competição global de mercado das empresas, o que envolve consideração não apenas ao próprio modelo de negócios, como o modelo de negócios companhias rivais, em um processo de interdependência estratégica. Como se pode perceber na Figura 15, uma dada companhia aérea ("Companhia Aérea 1") deve tomar a decisão entre "segmentar", ou "não segmentar" a cabine de suas aeronaves introduzindo uma ou mais classes econômicas premium, além da classe econômica tradicional. Simultaneamente a essa decisão, a companhia aérea deve definir o grau de densificação da cabine, inserindo mais ou menos fileiras de assentos, seja ao especificar novos pedidos ao fabricante, seja ao adaptar aeronaves usadas incorporadas à frota durante o processo de "fleet rollover" (substituição gradual da frota por aeronaves mais novas ou diferentes). Deve, portanto, decidir entre "adensar" ou "não adensar". Por sua vez, uma companhia aérea rival ("Companhia Aérea 2") deve tomar decisões similares, considerando a tomada de decisão esperada da concorrência. Havendo outras empresas na indústria, espera-se um posicionamento estratégico de mercado de cada uma delas, de forma a se atingir um equilíbrio com empresas voltadas aos modelos de negócio "Ultra Low Cost", "Low Cost" e "Mainline" (tradicionais).

As decisões tomadas por cada companhia aérea ao longo do jogo irá definir o seu posicionamento de preços junto aos consumidores, obtendo maior ou menor fatia de mercado junto aos passageiros mais e menos sensíveis à preço ("Low-yield PAX" e "High-yield PAX"). Esse posicionamento será decisivo na configuração do seu modelo de negócios. Assim, os ramos finais da árvore representam quatro possíveis posições estratégicas das companhias aéreas, cada uma resultante da combinação entre segmentar ou não segmentar a cabine e adensar ou não adensar os assentos. Por simplificação, o diagrama da Figura 15 exibe os resultados da Companhia Aérea 2, condicional ao fato da Companhia Aérea 1 optar pelo adensamento de cabine, mas outros resultados são possíveis. Assim, cada resultado corresponde a um tipo distinto de atuação no mercado:

- Quando a empresa não segmenta e adensa a cabine, o resultado é um posicionamento voltado para passageiros low-yield (baixo preço por quilômetro), isto é, muito sensíveis ao preço. Essa combinação leva a uma estrutura de custos reduzida e tende a caracterizar modelos Ultra Low Cost, cujo foco principal é maximizar capacidade e oferecer tarifas mínimas.

- Quando a empresa não segmenta e não adensa, ela opera com menor densidade e sem criar subprodutos dentro da classe econômica. Isso leva a um foco maior em passageiros "high-yield" (alto preço por quilômetro), mas aqueles menos sensíveis ao preço e que valorizam mais conforto relativo. Trata se de um posicionamento típico de companhias Mainline tradicionais.

- Quando a empresa segmenta e adensa, o resultado é um mix de passageiros low-yield e high-yield, pois a presença de um produto premium atrai consumidores com maior disposição a pagar, ao mesmo tempo em que o adensamento mantém custos baixos e permite captar clientes orientados a preço. Esse arranjo pode ser encontrado tanto em empresas Low Cost quanto em companhias Mainline com estratégias híbridas.

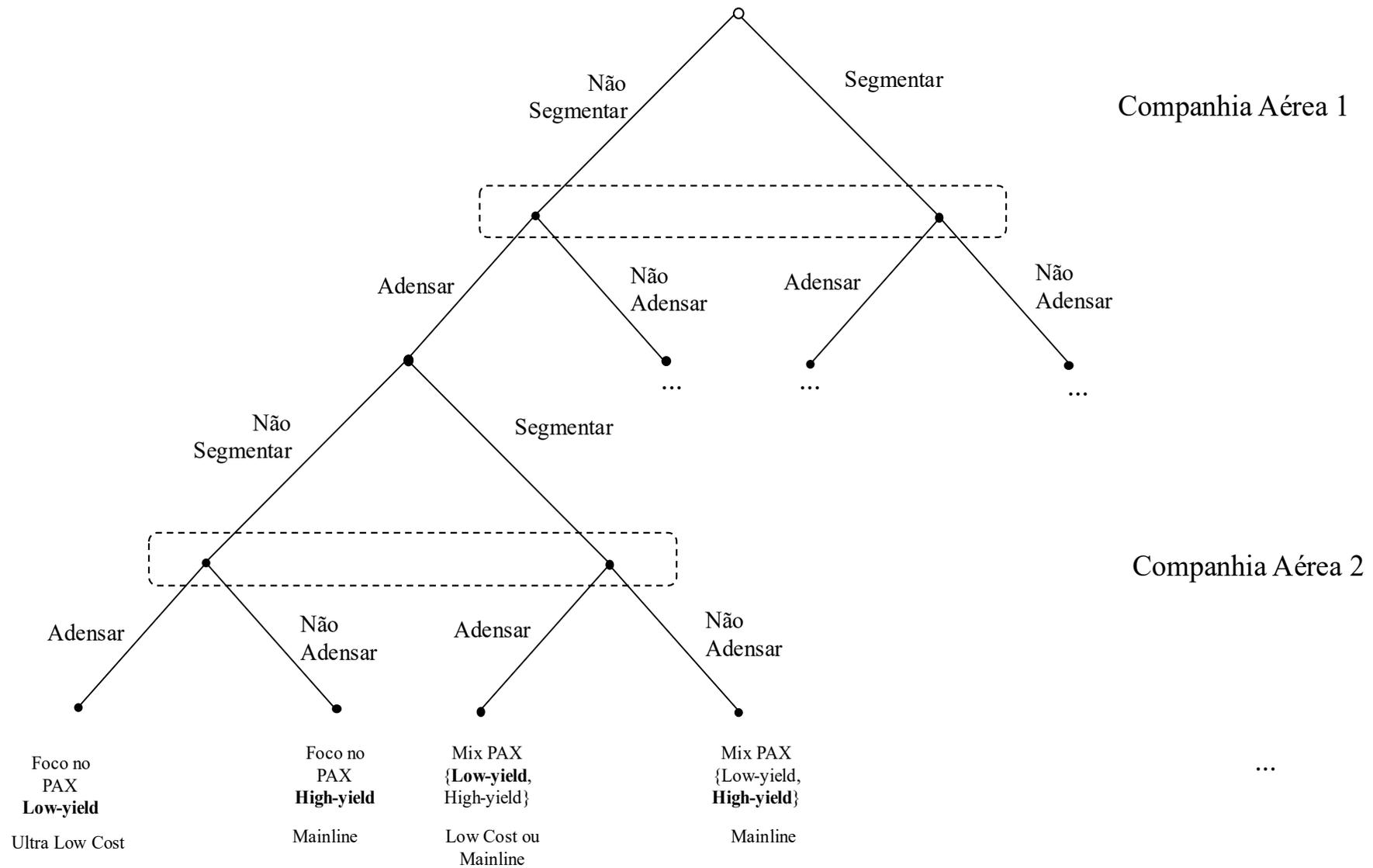

*Nota: O diagrama ilustra, de modo simplificado, um jogo sequencial entre duas companhias aéreas fictícias na escolha entre "Segmentar" ou "Não Segmentar", e "Adensar" ou "Não Adensar" a cabine, considerando aeronaves dotadas apenas de Classe Econômica. Os ramos finais ilustram os resultados finais da Companhia Aérea 2 condicionados à opção de adensamento da Companhia Aérea 1, embora outros desfechos possíveis não estejam representados. Essa estrutura resume a interação estratégica envolvida e as implicações dessas decisões para o posicionamento de mercado das empresas e a configuração de seu modelo de negócios, que pode ser negócio "Ultra Low Cost", "Low Cost" e "Mainline" (tradicional). "PAX Low-yield" são passageiros mais sensíveis ao preço, enquanto "PAX High-yield" são passageiros menos sensíveis ao preço e associados a tarifas mais altas*

**Figura 15 - Jogo decisório de configuração de cabine de aeronaves e posicionamento de mercado de companhias aéreas**

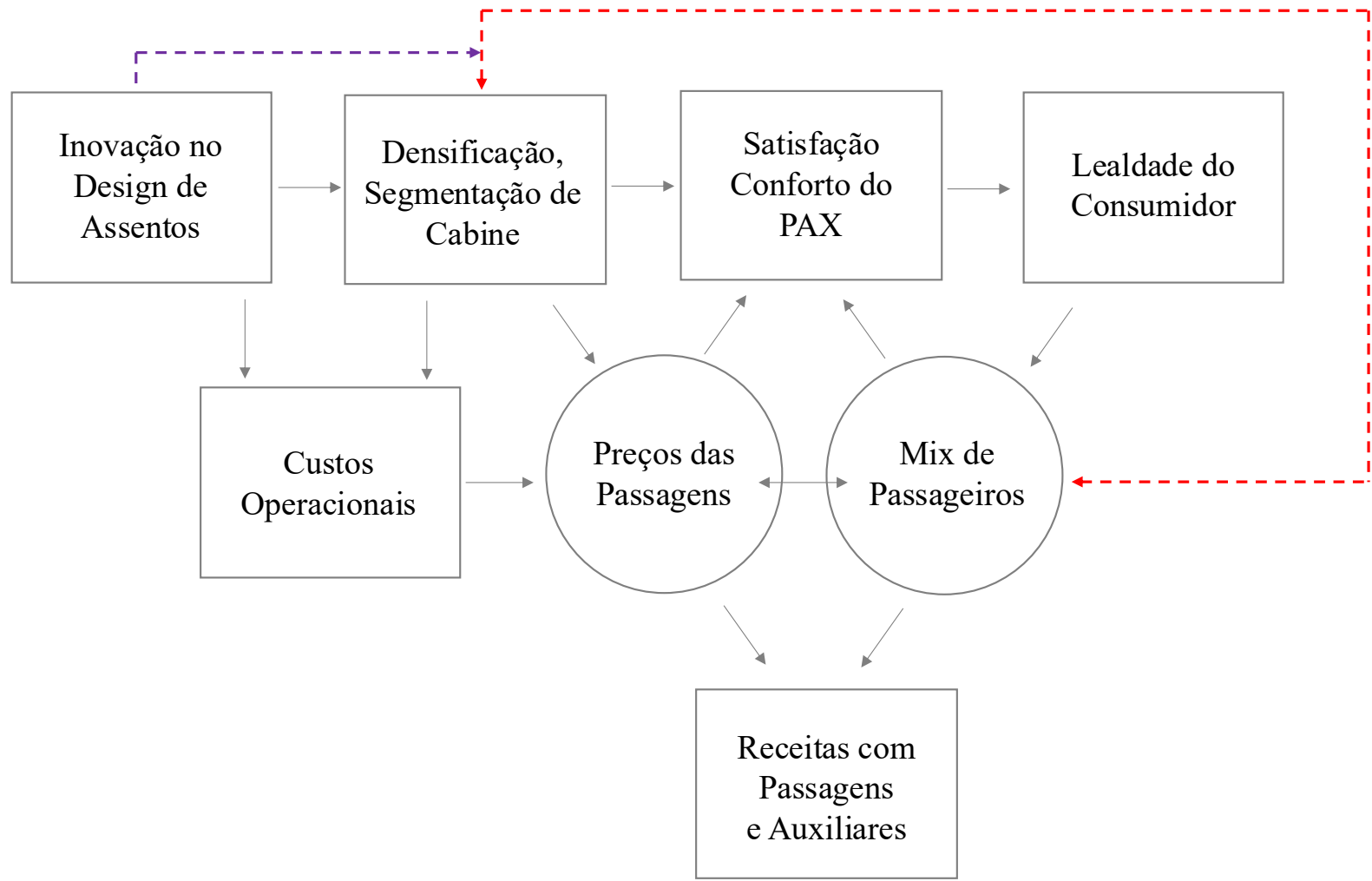

**Figura 16 - Modelo conceitual do impacto da introdução de design inovador de assentos**

- Quando a empresa segmenta e não adensa, ela oferece espaço relativamente maior somado a classes econômicas diferenciadas. Esse resultado também gera um mix de passageiros, mas com maior ênfase em consumidores high-yield, que valorizam conforto e serviços adicionais. Esse posicionamento tende a ser associado a empresas Mainline para capturar passageiros de maior valor.

Em conjunto, esses quatro desfechos do jogo da configuração de cabine sugerem como as escolhas sobre segmentação e densidade da cabine definem o público atendido, o modelo de negócio e o posicionamento competitivo de cada empresa na indústria aérea.

A Figura 16 apresenta o modelo conceitual proposto para analisar os impactos de mercado das decisões acerca da densidade de assentos em aeronaves e da introdução de design inovador de assentos na aviação comercial brasileira. O ponto de partida do modelo é a entrada de uma "Inovação no Design de Assentos", tratada como um choque externo que altera diretamente dois elementos da oferta: a "Densificação, Segmentação de Cabine" e os "Custos Operacionais". A densificação refere-se ao aumento ou redução do número de fileiras e ao espaço entre assentos, enquanto a segmentação diz respeito à criação de subprodutos dentro da classe econômica, como assentos premium, conforme o jogo discutido mais acima. Ao mesmo tempo, alterações nesses dois elementos afetam o comportamento dos preços no mercado, representados pelo bloco "Preços das Passagens". Esses preços são determinados por custos, pela disposição a pagar dos consumidores e pelo posicionamento competitivo das empresas. A inovação também exerce influência sobre a "Satisfação, Conforto do PAX", que por sua vez afeta diretamente a "Lealdade do Consumidor", elemento essencial para a manutenção de passageiros frequentes e para a sustentabilidade das receitas no tempo.

No centro do modelo conceitual da Figura 16 encontram-se os dois blocos circulares, "Preços das Passagens" e "Mix de Passageiros", cuja interação é endógena. Mudanças no preço atraem perfis distintos de passageiros, modificando o mix entre consumidores mais sensíveis ao preço e aqueles dispostos a pagar mais por conforto ou serviços adicionais. Essa composição afeta diretamente a receita total obtida com bilhetes e taxas extras, agrupadas no bloco "Receitas com Passagens e Auxiliares". A relação inversa também se mantém: alterações no "Mix de Passageiros" pressionam a estrutura de preços, exigindo reajustes para equilibrar custos, demanda e competitividade. Assim, temos que o modelo apresentado na Figura 16 evidencia como uma inovação inicialmente introduzida no interior da cabine desencadeia uma sequência de efeitos sobre estrutura de custos, precificação, composição da demanda, satisfação do passageiro e geração de receitas, configurando um ambiente competitivo dinâmico no qual pequenas mudanças de design podem gerar impactos sistêmicos expressivos.

A complexidade do jogo estratégico e do modelo conceitual apresentados evidencia que as decisões de configuração de cabine envolvem múltiplas interações simultâneas entre custos, preços, satisfação do passageiro, receitas auxiliares, lealdade e composição da demanda. Cada elemento retroalimenta os demais, formando um sistema no qual pequenas mudanças, como uma alteração no design de assentos, podem gerar efeitos amplificados em toda a estrutura competitiva do setor. Diante desse ambiente intrinsecamente interdependente, o presente trabalho concentra se apenas em uma parte específica desse conjunto de relações, focalizando o impacto das escolhas de layout sobre os "Preços das Passagens". Essa decisão analítica reconhece que os preços são endógenos ao "Mix de Passageiros", mas, neste primeiro momento, tratamos essa relação como exógena para viabilizar a identificação empírica. A avaliação plena da endogeneidade entre esses dois blocos é deixada para estudos futuros, que poderão explorar de forma mais profunda os mecanismos de retroalimentação entre perfil da demanda e estratégias de precificação.

Ao adotarmos esse recorte mais restrito, partimos da premissa de que as inovações em design de assentos se manifestam diretamente nas decisões de "Densificação, Segmentação de Cabine", como observado no estudo de Lee & Luengo Prado (2004), que analisou os efeitos de alterações de pitch e introdução de classes diferenciadas no início dos anos 2000. Situação semelhante ocorre com o fenômeno que investigamos, relacionado aos assentos conforto introduzidos em meados dos anos 2010, cujos impactos mais imediatos aparecem nas diferenças de preço entre produtos dentro da mesma aeronave. Assim, o modelo econométrico aqui desenvolvido é embebido na lógica do arcabouço conceitual apresentado, mas representa apenas uma pequena parcela de uma realidade mais ampla e complexa. Isso reforça a necessidade de interpretações cautelosas e de futuras extensões que possam capturar, de forma mais completa, a estrutura dinâmica que caracteriza o mercado de aviação comercial.

# V. Dados

A maioria dos dados utilizados no presente trabalho tem origem na pesquisa por questionários aplicados a passageiros em aeroportos brasileiros, contratada pela Empresa de Planejamento e Logística, EPL, em 2014. O relatório EPL (2014) apresenta os resultados oficiais dessa iniciativa, denominada "Pesquisa de Origem/Destino do Transporte Aéreo de Passageiros". O objetivo central da pesquisa foi ampliar o conhecimento sobre o fluxo de pessoas que utilizam o transporte aéreo no Brasil, tanto em viagens nacionais quanto internacionais, construindo um painel abrangente do comportamento e das características dos viajantes. A execução ficou a cargo do Instituto Olhar, Pesquisa e Informação Estratégica, responsável pela coleta e sistematização inicial das informações.

A pesquisa da EPL foi concebida para subsidiar a formulação do Plano Nacional de Logística Integrada, PNLI, instrumento voltado à proposição de ações de desenvolvimento e integração dos diversos modos de transporte do país. Segundo EPL (2014), o desenho amostral contemplou 65 aeroportos, que representavam 53,7% dos aeroportos com voos regulares e, simultaneamente, 99% de todos os passageiros embarcados no Brasil dois anos antes da coleta. Os questionários foram aplicados nas áreas de embarque nacional e internacional dos aeroportos selecionados, ao longo de quatro etapas distribuídas pelo ano de 2014 (janeiro/fevereiro, março/abril, maio e agosto), conforme descrito nos procedimentos metodológicos do estudo.

Para a presente pesquisa, a base original da EPL foi complementada com um conjunto de informações adicionais. Para viabilizar análises mais detalhadas sobre a localização dos passageiros a bordo, dezenas de milhares de cartões de embarque previamente digitalizados em formato de imagem para fins de auditoria da pesquisa, foram submetidos a um processo de transcrição e tabulação de dados referentes ao designador de assento do passageiro (como as poltronas de código 5A, 14F ou 22C), estruturando-as em formato adequado para análise. Esse processo permitiu identificar a localização exata dos assentos ocupados pelos passageiros e cruzar esses registros com as informações coletadas nas entrevistas com questionários.

Além disso, diversas outras bases da Agência Nacional de Aviação Civil (ANAC), foram integradas ao estudo, proporcionando maior profundidade e precisão às análises. Entre elas destacam-se: a Base de Dados Estatísticos do Transporte Aéreo, com dados agregados e microdados; os Microdados de Tarifas Aéreas Comercializadas; o banco Voo Regular Ativo (VRA), que reúne o histórico detalhado de voos; o Registro Aeronáutico Brasileiro (RAB), com informações sobre a frota nacional; e o Sistema de Registro de Operações (SIROS). A combinação dessas múltiplas fontes permitiu consolidar um banco detalhado para atender às necessidades das investigações propostas.

A Tabela 2 apresenta uma análise descritiva dos dados disponíveis. Ela contém um mapa de dispersão de passageiros (número e percentuais de entrevistados), segundo a localização do assento nas aeronaves. A tabela apresenta a distribuição dos passageiros entrevistados por fileira e por assento, permitindo observar como os viajantes se dispersam ao longo da cabine. O painel da esquerda mostra os números absolutos, o painel central expressa essas mesmas quantidades em percentuais por fileira e o painel da direita exibe a participação relativa de cada assento no total da amostra. Nota-se um padrão consistente de maior ocupação dos assentos de corredor, especialmente o assento D, que concentra parcelas superiores a 25 por cento em várias fileiras, seguido pelos assentos C e E. Os assentos da janela, A e F, têm participação intermediária, enquanto o assento do meio, B, apresenta participação menor na maioria das fileiras, reforçando a percepção de preferência reduzida por essa posição. Observa-se também um gradiente leve em direção ao centro da cabine, onde há maior concentração de entrevistados, sugerindo que, para aeronaves com embarque traseiro e dianteiro, a ocupação se distribui de forma relativamente equilibrada. Esses padrões ajudam a compreender como preferências estruturais dos passageiros podem influenciar a formação de preços e a demanda por determinados tipos de assento.

**Tabela 2 - Mapa de dispersão de passageiros (número e percentuais de entrevistados), segundo a localização do assento**

| | | Assentos | | | | | | Total |
|---|---|---|---|---|---|---|---|---|
| | | A | B | C | D | E | F | |
| Fileiras | 1 | 224 | 130 | 215 | 374 | 71 | 179 | 1,193 |
| | 2 | 527 | 321 | 388 | 550 | 168 | 331 | 2,285 |
| | 3 | 731 | 364 | 496 | 614 | 206 | 388 | 2,799 |
| | 4 | 712 | 364 | 487 | 660 | 216 | 426 | 2,865 |
| | 5 | 723 | 338 | 525 | 640 | 200 | 379 | 2,805 |
| | 6 | 717 | 436 | 611 | 698 | 192 | 408 | 3,062 |
| | 7 | 704 | 411 | 599 | 703 | 200 | 387 | 3,004 |
| | 8 | 694 | 406 | 606 | 694 | 173 | 371 | 2,944 |
| | 9 | 696 | 395 | 546 | 684 | 154 | 368 | 2,843 |
| | 10 | 692 | 356 | 566 | 624 | 168 | 363 | 2,769 |
| | 11 | 621 | 288 | 420 | 535 | 123 | 282 | 2,269 |
| | 12 | 634 | 303 | 388 | 594 | 126 | 302 | 2,347 |
| | 13 | 450 | 215 | 251 | 389 | 55 | 157 | 1,517 |
| | 14 | 582 | 328 | 399 | 538 | 144 | 316 | 2,307 |
| | 15 | 665 | 348 | 418 | 584 | 119 | 282 | 2,416 |
| | 16 | 578 | 311 | 381 | 520 | 86 | 269 | 2,145 |
| | 17 | 605 | 278 | 417 | 506 | 157 | 339 | 2,302 |
| | 18 | 580 | 246 | 369 | 468 | 157 | 372 | 2,192 |
| | 19 | 535 | 226 | 370 | 458 | 170 | 431 | 2,190 |
| | 20 | 560 | 224 | 380 | 418 | 178 | 437 | 2,197 |
| | 21 | 610 | 231 | 352 | 416 | 161 | 384 | 2,154 |
| | 22 | 550 | 213 | 351 | 433 | 155 | 381 | 2,083 |
| | 23 | 574 | 217 | 355 | 429 | 171 | 402 | 2,148 |
| | 24 | 433 | 203 | 308 | 388 | 121 | 318 | 1,771 |
| | 25 | 420 | 166 | 215 | 329 | 81 | 260 | 1,471 |
| | 26 | 385 | 167 | 226 | 296 | 96 | 261 | 1,431 |
| | 27 | 385 | 157 | 224 | 291 | 109 | 262 | 1,428 |
| | 28 | 360 | 131 | 176 | 249 | 76 | 251 | 1,243 |
| | 29 | 289 | 108 | 168 | 230 | 75 | 216 | 1,086 |
| | 30 | 186 | 65 | 126 | 146 | 51 | 144 | 718 |
| | 31 | 125 | 34 | 82 | 75 | 34 | 113 | 463 |
| | 32 | 85 | 31 | 54 | 40 | 18 | 93 | 321 |
| | Total | 16,632 | 8,011 | 11,469 | 14,573 | 4,211 | 9,872 | 64,768 |

| | | Assentos | | | | | | Total |
|---|---|---|---|---|---|---|---|---|
| | | A | B | C | D | E | F | |
| Fileiras | 1 | 19% | 11% | 18% | 31% | 6% | 15% | 100% |
| | 2 | 23% | 14% | 17% | 24% | 7% | 14% | 100% |
| | 3 | 26% | 13% | 18% | 22% | 7% | 14% | 100% |
| | 4 | 25% | 13% | 17% | 23% | 8% | 15% | 100% |
| | 5 | 26% | 12% | 19% | 23% | 7% | 14% | 100% |
| | 6 | 23% | 14% | 20% | 23% | 6% | 13% | 100% |
| | 7 | 23% | 14% | 20% | 23% | 7% | 13% | 100% |
| | 8 | 24% | 14% | 21% | 24% | 6% | 13% | 100% |
| | 9 | 24% | 14% | 19% | 24% | 5% | 13% | 100% |
| | 10 | 25% | 13% | 20% | 23% | 6% | 13% | 100% |
| | 11 | 27% | 13% | 19% | 24% | 5% | 12% | 100% |
| | 12 | 27% | 13% | 17% | 25% | 5% | 13% | 100% |
| | 13 | 30% | 14% | 17% | 26% | 4% | 10% | 100% |
| | 14 | 25% | 14% | 17% | 23% | 6% | 14% | 100% |
| | 15 | 28% | 14% | 17% | 24% | 5% | 12% | 100% |
| | 16 | 27% | 14% | 18% | 24% | 4% | 13% | 100% |
| | 17 | 26% | 12% | 18% | 22% | 7% | 15% | 100% |
| | 18 | 26% | 11% | 17% | 21% | 7% | 17% | 100% |
| | 19 | 24% | 10% | 17% | 21% | 8% | 20% | 100% |
| | 20 | 25% | 10% | 17% | 19% | 8% | 20% | 100% |
| | 21 | 28% | 11% | 16% | 19% | 7% | 18% | 100% |
| | 22 | 26% | 10% | 17% | 21% | 7% | 18% | 100% |
| | 23 | 27% | 10% | 17% | 20% | 7% | 19% | 100% |
| | 24 | 24% | 11% | 17% | 22% | 7% | 18% | 100% |
| | 25 | 29% | 11% | 15% | 22% | 6% | 18% | 100% |
| | 26 | 27% | 12% | 16% | 21% | 7% | 18% | 100% |
| | 27 | 27% | 11% | 16% | 20% | 8% | 18% | 100% |
| | 28 | 29% | 11% | 14% | 20% | 6% | 20% | 100% |
| | 29 | 27% | 10% | 15% | 21% | 7% | 20% | 100% |
| | 30 | 26% | 9% | 18% | 20% | 7% | 20% | 100% |
| | 31 | 27% | 7% | 18% | 16% | 7% | 24% | 100% |
| | 32 | 26% | 10% | 17% | 12% | 6% | 29% | 100% |
| | Total | 26% | 12% | 18% | 23% | 7% | 15% | 100% |

| | | Assentos | | | | | | Total |
|---|---|---|---|---|---|---|---|---|
| | | A | B | C | D | E | F | |
| Fileiras | 1 | 1% | 2% | 2% | 3% | 2% | 2% | 2% |
| | 2 | 3% | 4% | 3% | 4% | 4% | 3% | 4% |
| | 3 | 4% | 5% | 4% | 4% | 5% | 4% | 4% |
| | 4 | 4% | 5% | 4% | 5% | 5% | 4% | 4% |
| | 5 | 4% | 4% | 5% | 4% | 5% | 4% | 4% |
| | 6 | 4% | 5% | 5% | 5% | 5% | 4% | 5% |
| | 7 | 4% | 5% | 5% | 5% | 5% | 4% | 5% |
| | 8 | 4% | 5% | 5% | 5% | 4% | 4% | 5% |
| | 9 | 4% | 5% | 5% | 5% | 4% | 4% | 4% |
| | 10 | 4% | 4% | 5% | 4% | 4% | 4% | 4% |
| | 11 | 4% | 4% | 4% | 4% | 3% | 3% | 4% |
| | 12 | 4% | 4% | 3% | 4% | 3% | 3% | 4% |
| | 13 | 3% | 3% | 2% | 3% | 1% | 2% | 2% |
| | 14 | 3% | 4% | 3% | 4% | 3% | 3% | 4% |
| | 15 | 4% | 4% | 4% | 4% | 3% | 3% | 4% |
| | 16 | 3% | 4% | 3% | 4% | 2% | 3% | 3% |
| | 17 | 4% | 3% | 4% | 3% | 4% | 3% | 4% |
| | 18 | 3% | 3% | 3% | 3% | 4% | 4% | 3% |
| | 19 | 3% | 3% | 3% | 3% | 4% | 4% | 3% |
| | 20 | 3% | 3% | 3% | 3% | 4% | 4% | 3% |
| | 21 | 4% | 3% | 3% | 3% | 4% | 4% | 3% |
| | 22 | 3% | 3% | 3% | 3% | 4% | 4% | 3% |
| | 23 | 3% | 3% | 3% | 3% | 4% | 4% | 3% |
| | 24 | 3% | 3% | 3% | 3% | 3% | 3% | 3% |
| | 25 | 3% | 2% | 2% | 2% | 2% | 3% | 2% |
| | 26 | 2% | 2% | 2% | 2% | 2% | 3% | 2% |
| | 27 | 2% | 2% | 2% | 2% | 3% | 3% | 2% |
| | 28 | 2% | 2% | 2% | 2% | 2% | 3% | 2% |
| | 29 | 2% | 1% | 1% | 2% | 2% | 2% | 2% |
| | 30 | 1% | 1% | 1% | 1% | 1% | 1% | 1% |
| | 31 | 1% | 0% | 1% | 1% | 1% | 1% | 1% |
| | 32 | 1% | 0% | 0% | 0% | 0% | 1% | 0% |
| | Total | 100% | 100% | 100% | 100% | 100% | 100% | 100% |

*Notas: Análise dos dados da Pesquisa de Origem/Destino do Transporte Aéreo de Passageiros (EPL, 2014), versão compilada. A tabela não representa a configuração de uma aeronave, mas apenas permite a visualização da dispersão espacial, considerando diferentes tipos de aeronave e configurações. As cores representam a intensidade relativa de ocupação: tons de azul indicam menores concentrações de passageiros e tons de vermelho indicam maiores concentrações, facilitando a identificação visual dos padrões de preferência por assento e da distribuição ao longo das fileiras. Dos 122 mil passageiros entrevistados, foram selecionados apenas aqueles com voos domésticos das quatro maiores empresas aéreas com voos identificados. Questionários com erros ou omissões, fotografias de cartões de embarque ilegíveis ou cortados, e fileiras acima da de número 32 foram desconsiderados.*

# VI. METODOLOGIA

O presente estudo visa estimar o efeito da densidade de assentos, bem como da localização de assentos dentro do layout de cabine, sobre o preço das passagens aéreas. A motivação empírica dialoga com evidências de que preferências relacionadas ao espaço e à seleção de assentos influenciam padrões de tarifação. Como visto, Rouncivell, Timmis e Ison (2018) mostram que a sensibilidade ao preço apresenta relação negativa com a disposição declarada a pagar pela marcação de assento, indicando que decisões de compra incorporam trade-offs entre custo e conveniência. Já Lee e Luengo Prado (2004) documentam que ajustes no pitch afetam os preços das passagens, ainda que de forma diferenciada entre estratégias de aumento uniforme ou restrito a fileiras específicas. Esses trabalhos prévios ilustram como atributos físicos da cabine e políticas de diferenciação podem produzir variações sistemáticas nos preços observados, fundamento que orienta a especificação empírica adotada nesta análise.

Como visto, a base de dados é constituída pela Pesquisa EPL, Estudo de Perfil do Passageiro, com entrevistas presenciais em aeroportos brasileiros e coleta de cartões de embarque. A amostra utilizada nas estimações contém 15.517 observações após limpeza e pareamento (em algumas especificações mais simples do modelo, 15.634), concentrada em voos das companhias Gol e TAM (atual Latam).

O conjunto de variáveis utilizadas na modelagem econométrica é apresentado a seguir. Para efeito das estimações, todas as variáveis contínuas foram transformadas em logaritmo.

- P é a variável dependente dos modelos, sendo o preço pago pela passagem aérea, conforme declarado pelo passageiro nos questionários da Pesquisa EPL (fonte: questionários da Pesquisa EPL). A variável é expressa em unidades monetárias correntes (R$).
- ADV é a antecedência de compra em dias, informada pelo passageiro na Pesquisa EPL (fonte: questionários da Pesquisa EPL). A variável mede quantos dias se passaram entre a compra e o voo, permitindo observar como o momento da compra se relaciona com o preço.
- DIST é a distância entre os aeroportos da etapa, em quilômetros, obtida de registros estatísticos oficiais (fonte: Dados Estatísticos do Transporte Aéreo da ANAC, cálculos próprios). A variável está ligada aos custos do voo e à possível concorrência com modos terrestres de transporte.
- BSN é a indicação de viagem a negócios, declarada diretamente pelo passageiro na Pesquisa EPL (fonte: questionários da Pesquisa EPL). A variável distingue perfis de viajantes e visa capturar possíveis diferenças de sensibilidade a preço que se manifestam via preço pago das passagens.
- FLTIME é o tempo de voo em minutos, obtido de dados operacionais oficiais (fonte: Dados Estatísticos do Transporte Aéreo da ANAC, cálculos próprios). A variável está associada aos custos do voo, à concorrência com o transporte terrestre e a desconfortos ou custos de tempo durante a viagem.
- SHIPMENT é o volume total transportado no porão, em quilogramas, incluindo bagagens e carga (fonte: Dados Estatísticos do Transporte Aéreo da ANAC, cálculos próprios). A variável representa o uso do porão e se relaciona com custos e tempo de processamento ligados ao turnaround time da aeronave.
- REVPAX é o número de passageiros pagos a bordo, obtido de dados estatísticos oficiais (fonte: Dados Estatísticos do Transporte Aéreo da ANAC, cálculos próprios). A variável indica a densidade de tráfego da rota e seus efeitos operacionais.
- LF é o fator de ocupação, definido como o percentual de assentos oferecidos na aeronave que foram ocupados por passageiros pagantes com base em registros oficiais (fonte: Dados Estatísticos do Transporte Aéreo da ANAC, cálculos próprios). A variável mostra o nível de ocupação e a escassez de assentos.
- FUELP é o preço do combustível na região dos aeroportos da rota, expresso em em unidades monetárias correntes (R$), obtido de registros consolidados (fonte: Agência Nacional do Petróleo, ANP, cálculos próprios). A variável representa o componente de custo associado ao insumo energético usado no voo.
- HUB é uma dummy indicadora de que o passageiro realizou ao menos uma conexão, informação declarada na Pesquisa EPL (fonte: questionários da Pesquisa EPL). A variável reflete condições de rede e acessibilidade do aeroporto.
- SEATSH é a participação da companhia aérea no total de assentos oferecidos na rota naquele dia, obtida de dados de oferta (fonte: Dados Estatísticos do Transporte Aéreo da ANAC, cálculos próprios), multiplicada por 100. A variável indica a posição competitiva da empresa no mercado da rota.

- RHHI é o índice de concentração da rota, calculado a partir da participação de assentos das empresas (fonte: Dados Estatísticos do Transporte Aéreo da ANAC, cálculos próprios). A variável mostra o nível de competição entre as companhias que operam a rota.
- LASTROW é uma dummy indicadora de que o assento do passageiro fica na última fileira, combinando o assento informado na Pesquisa EPL com mapas de cabine (fonte: cartões de embarque digitalizados da Pesquisa EPL, guias Panrotas). A variável representa uma posição menos desejada dentro da cabine.
- EMERGEXIT é uma dummy indicadora de assento situado na saída de emergência, obtida cruzando o assento declarado na Pesquisa EPL com mapas de cabine (fonte: cartões de embarque digitalizados da Pesquisa EPL, guias Panrotas). A variável representa locais com maior espaço para as pernas e regras de uso específicas.
- COMFORT é uma dummy indicadora de assento de pitch ampliado das companhias aéreas, determinada pelo assento constante no cartão de embarque da Pesquisa EPL e por sua posição no mapa de cabine (fonte: cartões de embarque digitalizados da Pesquisa EPL, guias Panrotas). Assento de pitch ampliado é uma posição na cabine cujo espaçamento entre fileiras é maior que o padrão do restante da aeronave, oferecendo conforto adicional dentro da própria classe econômica. Esses assentos foram identificados em aeronaves Boeing 737 da Gol e em alguns dos modelos Airbus A319, A320 e A321 da TAM (atual Latam), a partir de análise dos mapas de assentos.
- COMFORT (placebo) é uma dummy que identifica assentos equivalentes aos de pitch ampliado, mas presentes em aeronaves do mesmo modelo e da mesma companhia aérea que não ofereciam essa configuração interna (fonte: cartões de embarque digitalizados da Pesquisa EPL, guias Panrotas). Essa variável serve como controle para verificar se eventuais efeitos atribuídos ao conforto a bordo, medido por COMFORT decorrem de erro de classificação ou de características estruturais do layout que não estão relacionadas ao produto diferenciado.
- MIDDLE é uma dummy indicadora de assento do meio, determinada a partir do assento informado na Pesquisa EPL e confirmada pelos mapas de cabine (fonte: cartões de embarque digitalizados da Pesquisa EPL, guias Panrotas). A variável representa uma posição mais confinada lateralmente.
- MIDDLE × ADV são interações entre a variável MIDDLE e faixas de antecedência classificadas como 1w, 2w, 3w e mais de 3w, onde w corresponde a semanas (fonte: cartões de embarque digitalizados da Pesquisa EPL, guias Panrotas; questionários da Pesquisa EPL, cômputos próprios). As variáveis mostram como o efeito do assento do meio muda conforme o momento da compra.
- IROWDENS, é um indice de densidade de fileiras. É a razão entre o total de fileiras instaladas na aeronave e o valor máximo de fileiras documentado internacionalmente para o mesmo modelo (fonte: Guia Panrotas e website Seatguru[19]). O índice representa uma proxy para a compactação estrutural da cabine, indicando maior intensidade de uso do espaço, redução do conforto, e potencial redução do custo por assento quando seus valores são mais altos. O índice é normalizado para 100.
- IPITCH é um índice de seat pitch. É um índice de seat pitch. É a razão entre o espaçamento médio real entre as fileiras da aeronave, medido em polegadas, e o maior valor de pitch documentado internacionalmente para o mesmo modelo modelo (fonte: Guia Panrotas e website Seatguru[20]). O cálculo utiliza um pitch médio ponderado pela distribuição de fileiras nas diferentes seções da cabine. O índice reflete uma proxy para o conforto longitudinal oferecido pela configuração de assentos, aumentando quando o espaço por passageiro é maior. O índice é normalizado para 100.

Para a estimação da modelagem econométrica proposta, foi utilizada a abordagem Post-Double-Selection LASSO (PDS-LASSO), adequada a fenômenos que requerem o uso de muitos controles potenciais. A principal motivação para o uso desse procedimento está no risco de viés por variáveis omitidas associado à complexidade da formação de preços das passagens aéreas. Situações simples ilustram esse problema, como diferenças de demanda entre horários mais disputados e horários menos procurados, variações específicas de determinados dias que influenciam simultaneamente o fluxo de passageiros e o nível das tarifas, ou características individuais relacionadas ao perfil do viajante. Por exemplo, passageiros mais velhos podem planejar com maior antecedência, enquanto passageiros mais jovens podem comprar mais perto da data do voo, e indivíduos com maior renda ou alta frequência de viagens podem ter padrões de compra distintos dos demais. Para reduzir esses vieses potenciais, a estratégia de estimação adota um conjunto amplo de controles candidatos que são submetidos à penalização do LASSO e retidos apenas quando apresentam relação estatística relevante. Em seguida, os regressores de interesse são estimados por Mínimos Quadrados Ordinários, adicionando-se os controles previamente selecionados. Os erros-padrão estimados foram robustos ao agrupamento por pares de aeroportos.

---

[19] Website seatguru.com, conforme pesquisa em Agosto de 2021.
[20] idem.

No que tange aos controles penalizados pelo PDS-LASSO, foram utilizadas dummies de datas da pesquisa (130) e de horários de partida do voo (24). Em especificações de alta dimensão mais profundas, foram incluídos ainda controles relacionados aos aeroportos envolvidos (55) e um conjunto extenso de controles de perfil de passageiros (1761) oriundos da pesquisa EPL. Esses últimos combinam faixas etárias, faixas de renda, sexo e frequência de viagens no último ano, o que permite representar, por exemplo, grupos de passageiros mais jovens ou mais velhos, de renda mais baixa ou mais alta, homens ou mulheres, e viajantes ocasionais ou frequentes. Cada combinação gera uma dummy específica de perfil individual que entra como candidata no processo de penalização, contribuindo para absorver componentes não observados do comportamento de compra que possam influenciar simultaneamente o preço pago e a posição do passageiro na cabine.

## VII. Resultados das estimações

Os resultados das estimações estão apresentados na Tabela 3. A tabela utiliza oito colunas para mostrar como as estimativas mudam conforme as variáveis regressoras são introduzidas de forma cumulativa, partindo da especificação mais simples à esquerda e avançando para especificações mais completas à direita. A Coluna (5) corresponde à especificação central para interpretação. As Colunas (6) a (8) oferecem variações dessa mesma estrutura, seja por meio da inclusão de interações com MIDDLE na Coluna (6), da incorporação de controles de alta dimensão de aeroportos e características de passageiros na Coluna (7), ou da substituição de IROWDEN pela proxy IPITCH na Coluna (8).

ADV é negativo e estatisticamente significante em todas as especificações. Isso indica associação entre maior antecedência de compra e tarifas menores, o que é compatível com padrões conhecidos de algoritmos de gerenciamento de receita. DIST é positivo em todas as colunas e permanece estatisticamente significante mesmo após a inclusão de FLTIME, sugerindo que distância e duração do voo captam componentes distintos dos custos operacionais não observáveis. Na Coluna (1), sem FLTIME, DIST absorve a maior parte dessa variação. Quando FLTIME passa a integrar o modelo a partir da Coluna (3) até a (8), o coeficiente de DIST diminui, indicando que ambos passam a compartilhar a explicação estatística desse componente de custo, com possível sobreposição parcial entre as duas medidas.

BSN é positivo e estatisticamente significante nas Colunas (2) a (8), com pouca variação entre especificações, o que sinaliza estabilidade do coeficiente estimado. Isso sugere que a diferença de tarifa para passageiros a negócios não é sensível ao conjunto de controles incluídos nas diferentes especificações e não parece resultar de omissão sistemática de variáveis nesta amostra. Esse comportamento é compatível com práticas de diferenciação tarifária das companhias aéreas, que com esquemas de segmentação podem parcialmente conferir um premium de preços advindo de viajantes a negócios e em rotas mais fortemente marcadas pela predominância desses passageiros no mix de consumidores.

Nas outras proxies relacionadas a aspectos operacionais, estratégicos e de mercado, os sinais também permanecem relativamente consistentes entre as colunas. FUELP é positivo e estatisticamente significante em todas as especificações, o que sugere repasse parcial de choques de combustível. HUB apresenta coeficiente positivo estável, compatível com características de rede e possíveis diferenças de poder de mercado nos aeroportos concentradores. RHHI é positivo e estatisticamente significante sempre que incluído, com valores maiores nas Colunas (7) e (8), indicando que, com controles mais detalhados de heterogeneidade de passageiros, a associação entre concentração e tarifa se torna mais pronunciada. Essa interpretação deve ser vista com cautela, dada a possibilidade de variáveis omitidas e limitações de identificação.

Nas variáveis de capacidade e ocupação, observa-se que os coeficientes seguem o padrão esperado, mas perdem força conforme os controles ficam mais granulares. LF é positivo e estatisticamente significante nas Colunas (3) a (6) e deixa de ser estatisticamente significante nas Colunas (7) e (8), indicando que, após controlar por perfis de passageiro e aeroportos, o fator de ocupação residual explica menos variação de tarifas. REVPAX é negativo e estatisticamente significante até a Coluna (6) e enfraquece depois, indicando que proxies a densidade de tráfego a bordo captura a composição tarifária nas especificações em que os efeitos fixos entram em maior número. SEATSH também é positivo e estatisticamente significante nas Colunas (3) a (6), perdendo significância nas Colunas (7) e (8), compatível com o fato de que parte da informação que ele carrega é absorvida pelos efeitos de aeroporto e características do passageiro. O mesmo ocorre com SHIPMENT.

A interpretação das variáveis de densidade e localização de assentos (LASTROW, EMERGEXIT COMFORT, MIDDLE, IROWDENS e IPITCH) deve começar pelo fato de que, no período analisado, ainda não havia uma sistemática de cobrança de tarifas de seleção de assentos. Esse era um momento em que essa prática estava em processo de introdução pela indústria, especialmente nas fileiras dianteiras associadas aos assentos de pitch ampliado. Assim, qualquer coeficiente estimado para essas variáveis reflete apenas diferenças no preço pago pelos passageiros que efetivamente ocuparam cada tipo de assento, e não o efeito de uma tarifa específica. Um coeficiente estatisticamente significante indica somente que, em média, quem ocupou aquele assento pagou um valor maior ou menor, condicionais aos controles do modelo. Essa análise permite entender as possibilidades de cobrança de tarifas de seleção para reordenamento de passageiros a bordo visando uma maior eficiência alocativa. Os resultados obtidos sugerem alguns padrões interessantes, mas a variabilidade observada e as limitações na abordagem e no estimador indicam que o tema ainda exige investigação adicional para esclarecer melhor os mecanismos envolvidos.

IROWDENS é negativo e estatisticamente significante nas Colunas (1) a (6) e permanece negativo na Coluna (7), com coeficiente e significância estatística menores nessa especificação. Isso indica que maior densidade de assentos possivelmente está associada a preços mais baixos, o que é consistente com a ideia de que aumentar a quantidade de fileiras reduz o custo por assento. Esses resultados sugerem a presença de economias ligadas ao tamanho efetivo da cabine no processo de precificação. Na Coluna (8), utilizamos outra forma de medir esse mesmo mecanismo, IPITCH, que substitui IROWDENS e apresenta coeficiente positivo e estatisticamente significante. Nesse caso, maior espaçamento médio entre fileiras aparece associado a preços mais altos, refletindo o mesmo princípio a partir de outra métrica. A troca de indicador mantém a interpretação econômica e altera minimamente os critérios de ajuste do modelo. Por serem indicadores construídos a partir de referências externas de densidade ou de espaçamento adotadas pela indústria em diferentes operadores e países, medidas como IROWDENS e IPITCH refletem características estruturais da configuração da aeronave, e não percepções diretas do passageiro. O passageiro observa apenas o espaço disponível no seu próprio assento, sem referência às densidades relativas utilizadas por outras companhias ou pelos fabricantes. Por isso, esses indicadores tendem a capturar componentes não observáveis relacionados a custos e à configuração técnica da cabine, e não preferências explícitas dos consumidores. Por essa razão, os sinais negativos em IROWDENS (Colunas 1 a 7) e o sinal positivo em IPITCH (Coluna 8) não devem ser interpretados como evidência de maior ou menor disposição a pagar por parte dos passageiros no que se refere ao layout de cabine. Esses coeficientes refletem principalmente variações estruturais da configuração da aeronave e componentes de custo associados ao uso do espaço, e não diferenças diretamente percebidas ou valorizadas pelo usuário no momento da compra.

Os atributos locacionais de cabine medidos por variáveis dummy (LASTROW, EMERGEXIT, COMFORT e MIDDLE) apresentam padrões que, em alguns casos, diferem do que seria esperado à luz das práticas atuais. Um coeficiente estatisticamente significante dessas variáveis indica que, em média, quem ocupou aquele assento pagou um preço maior ou menor, condicionais aos controles do modelo. Entretanto, LASTROW não é estatisticamente significante, o que indica ausência de diferença sistemática no preço pago pelos passageiros alocados à última fileira, após controlar pelos demais fatores. EMERGEXIT também não é estatisticamente significante, mesmo sendo uma posição que hoje costuma ter cobrança específica. Isso sugere que, naquele período, essa localização não gerava variações detectáveis no preço pago pelos ocupantes desses assentos. COMFORT apresenta coeficientes positivos, mas estatisticamente significantes apenas ao nível de 10 por cento nas Colunas (5) a (8). Isso indica evidência limitada de que passageiros sentados em assentos com pitch ampliado pagavam preços de passagens ligeiramente maiores. Essa interpretação é compatível com a presença de grupos com maior disposição a pagar. O placebo de COMFORT não é estatisticamente significante, o que sugere que, ao comparar aeronaves do mesmo modelo e da mesma companhia que não possuíam assentos com pitch ampliado, aparentemente não estamos lidando com um efeito espúrio. Isso reduz a possibilidade de que o coeficiente observado em COMFORT resulte de erro de medição ou de fatores estruturais não relacionados ao assento em si. Entretanto, a ausência de significância em níveis estatísticos amplamente aceitos indica que os resultados de COMFORT não se diferenciam de forma clara do placebo, o que sugere que, nesse período, a oferta de espaço extra não desempenhava papel de segmentação efetiva no preço pago pelos passageiros.

MIDDLE apresenta um padrão que requer análise mais pormenorizada. Na Coluna (5), o coeficiente estimado dessa dummy é positivo e estatisticamente significante, um resultado contraintuitivo, mas conceitualmente possível no contexto analisado, que, conforme discutido, corresponde a um período anterior à cobrança de tarifas de marcação de assento. Nesse cenário, passageiros com maior disposição a pagar pela passagem frequentemente chegavam ao booking com baixa antecedência, encontrando as posições laterais já preenchidas e restando majoritariamente assentos centrais. Esse comportamento revela os incentivos econômicos que as companhias aéreas têm para introduzir tarifas de seleção de assento, como forma de fortalecer a eficácia de seus esquemas de segmentação, no sentido de direcionar os melhores assentos para passageiros com maior disposição a pagar e maior fidelidade, que tendem a valorizar mais os atributos do serviço. Isso fornece evidências que aparentemente permitem justificar a racionalidade econômica desse tipo de cobrança na indústria.

As interações MIDDLE × ADV, apresentadas na Coluna (6) e nas demais, esclarecem melhor o mecanismo que os dados apontam. As interações indicam que os passageiros que pagam mais caro e acabam sentando no assento do meio são possivelmente aqueles que compram com antecedência muito baixa, especialmente dentro de uma semana do voo. Para ADV de uma semana (1w), o efeito condicional estimado é positivo e estatisticamente significante. Para duas semanas (2w), o efeito ainda é positivo, porém menor e estatisticamente significante apenas a 10%. Para 3 semanas (3w), o efeito se dissipa. Para mais de 3 semanas (>3w), o coeficiente se torna negativo e estatisticamente significante também apenas ao nível de 10%. Ou seja, os resultados proporcionam evidências de que quem chegava mais tarde ao booking encontrava menos opções de assentos, e a maior ocupação fazia com que sobrassem principalmente assentos do meio. O aparente prêmio de passagens associado a esse tipo de assento, portanto, não reflete preferência, mas sim a dinâmica de compra tardia em um período sem cobrança de marcação de assento.

Critérios de ajuste e escolha de especificação indicam preferência pelas colunas mais completas. AIC e RMSE diminuem de forma monotônica conforme entram mais controles e atributos de cabine. As Colunas (6) e (7) combinam completude com manutenção de IROWDENS e apresentam baixos valores de AIC e RMSE, sendo adequadas quando o objetivo é mensurar diretamente a relação entre densidade de fileiras e preços. A Coluna (8) apresenta o menor AIC e o menor RMSE, ao custo de trocar a métrica de densidade por espaçamento médio, funcionando como teste de robustez e como interpretação em termos de disposição a pagar.

Em síntese, três pontos estruturais se destacam. Primeiro, a densidade da cabine importa para a formação de preços, seja medida por IROWDENS, seja por IPITCH, com sinais opostos porém coerentes e compatíveis com economias de escala associadas ao tamanho da aeronave medido em número de fileiras de assentos. Esse resultado dialoga diretamente com a evidência de Lee & Luengo Prado (2004) de que maior espaço por passageiro pode, em certos contextos, produzir diferenças tarifárias, embora em nosso caso a granularidade dos microdados permita identificar efeitos internos à aeronave que o estudo deles não observou. Segundo, o assento do meio não apresenta prêmio negativo próprio como seria de se esperar, e seu efeito decorre do padrão de compras tardias, como evidenciado pelas interações com ADV, indicando que, à época, passageiros com perfis de maior disposição a pagar frequentemente acabavam em posições menos desejáveis na cabine. Esse achado amplia a interpretação da literatura ao revelar que preferências individuais podem ser mascaradas por mecanismos de alocação, algo não captado pelo desenho antes e depois de mudanças estruturais de pitch analisado por aqueles autores. Terceiro, variáveis vinculadas ao contexto competitivo e de rede, como RHHI e HUB, elevam preços, enquanto indicadores agregados como LF e REVPAX perdem relevância quando o modelo passa a incluir controles finos de data, aeroporto e perfil, reforçando que a heterogeneidade intra aeronave adiciona uma camada de variação que não está presente em estudos baseados em médias de mercado. Por fim, os resultados indicam que o período amostral condiciona de forma importante essas relações. Como se trata de uma fase anterior ao uso sistemático de instrumentos de diferenciação associados à localização de assentos, os padrões observados sugerem eficácia limitada dos mecanismos de segmentação então existentes. Isso aponta para um espaço de aprimoramento que seria explorado nos anos seguintes, com a adoção gradual de práticas voltadas à distinção entre produtos e à alocação mais eficiente de passageiros conforme suas preferências e disposição a pagar.

**Tabela 3 – Resultados das estimações e checks de robustez**

|  | (1) | (2) | (3) | (4) | (5) | (6) | (7) | (8) |
|---|---|---|---|---|---|---|---|---|
| ADV | -0.0044*** | -0.0031*** | -0.0032*** | -0.0032*** | -0.0033*** | -0.0029*** | -0.0030*** | -0.0030*** |
| DIST | 0.5320*** | 0.5516*** | 0.2626*** | 0.2362*** | 0.2385*** | 0.2390*** | 0.3151*** | 0.3090*** |
| BSN |  | 0.3993*** | 0.3941*** | 0.3868*** | 0.3912*** | 0.3827*** | 0.3828*** | 0.3829*** |
| FLTIME |  |  | 0.4401*** | 0.4824*** | 0.4776*** | 0.4790*** | 0.3018*** | 0.3056*** |
| SHIPMENT |  |  | 0.0382*** | 0.0356** | 0.0376*** | 0.0382*** | 0.0294* | 0.0220 |
| REVPAX |  |  | -0.2849*** | -0.1725*** | -0.1698*** | -0.1694*** | -0.1034* | -0.0893 |
| LF |  |  | 0.2648*** | 0.1766*** | 0.1632*** | 0.1575*** | 0.0722 | 0.0609 |
| FUELP |  |  | 1.0256*** | 0.7828*** | 0.7424*** | 0.7554*** | 0.8568*** | 0.7968*** |
| HUB |  |  | 0.5302*** | 0.5316*** | 0.5306*** | 0.5363*** | 0.5025*** | 0.5043*** |
| SEATSH |  |  |  | 0.0675*** | 0.0668*** | 0.0674*** | 0.0274 | 0.0235 |
| RHHI |  |  |  | 0.1457*** | 0.1490*** | 0.1477*** | 0.1887*** | 0.2011*** |
| LASTROW |  |  |  |  | 0.0322 | 0.0370 | 0.0503 | 0.0455 |
| EMERGEXIT |  |  |  |  | 0.0306 | 0.0268 | 0.0376 | 0.0455 |
| COMFORT |  |  |  |  | 0.6669* | 0.6602* | 0.6512* | 0.6600* |
| COMFORT (placebo) |  |  |  |  |  | 0.2207 | 0.1644 | 0.1816 |
| MIDDLE |  |  |  |  | 0.0649*** |  |  |  |
| MIDDLE × ADV (1w) |  |  |  |  |  | 0.2949*** | 0.2888*** | 0.2881*** |
| MIDDLE × ADV (2w) |  |  |  |  |  | 0.1018** | 0.0869* | 0.0854* |
| MIDDLE × ADV (3w) |  |  |  |  |  | -0.0232 | -0.0326 | -0.0347 |
| MIDDLE × ADV (>3w) |  |  |  |  |  | -0.0358* | -0.0367* | -0.0359* |
| IROWDENS | -0.6120*** | -0.5464*** | -0.6761*** | -0.7222*** | -0.6872*** | -0.7293*** | -0.4062** |  |
| IPITCH |  |  |  |  |  |  |  | 1.0998*** |
| Estimator | PDS/LASSO | PDS/LASSO | PDS/LASSO | PDS/LASSO | PDS/LASSO | PDS/LASSO | PDS/LASSO | PDS/LASSO |
| Airport-Pair Clusters | 333 | 333 | 333 | 333 | 333 | 333 | 333 | 333 |
| Svy Date Controls | 1/130 | 4/130 | 51/130 | 53/130 | 60/130 | 61/130 | 45/130 | 45/130 |
| Flight Time Controls | 3/24 | 3/24 | 4/24 | 4/24 | 4/24 | 4/24 | 2/24 | 2/24 |
| Airport Controls | No | No | No | No | No | No | 19/55 | 21/55 |
| PAX Profile Controls | No | No | No | No | No | No | 230/1761 | 230/1761 |
| AIC Statistic | 33,398 | 32,230 | 31,409 | 31,287 | 31,271 | 31,154 | 30,819 | 30,792 |
| BIC Statistic | 33,460 | 32,322 | 31,913 | 31,822 | 31,890 | 31,812 | 32,387 | 32,368 |
| Adj R2 Statistic | 0.2111 | 0.2681 | 0.2972 | 0.3029 | 0.3041 | 0.3096 | 0.3245 | 0.3256 |
| RMSE Statistic | 0.7039 | 0.6780 | 0.6643 | 0.6616 | 0.6611 | 0.6585 | 0.6513 | 0.6508 |
| Nr Observations | 15,634 | 15,634 | 15,517 | 15,517 | 15,517 | 15,517 | 15,517 | 15,517 |

*Notas: Resultados da estimativa nas colunas (1)-(8) produzidos pela metodologia baseada em LASSO pós-dupla seleção de Belloni et al. (2012, 2014a,b) (PDS-LASSO). Erros padrão robustos à heteroscedasticidade. Estimativas das variáveis de controle omitidas. Todas as variáveis de controle são penalizadas pelo LASSO. Representações do valor P: \*\*\*p<0,01, \*\*p<0,05, \*p<0,10.*

## VIII. Conclusões

O presente trabalho investigou a relação entre o layout da cabine e os preços das passagens aéreas pagos pelos viajantes em voos domésticos, a partir de dados de uma pesquisa com questionários realizada em aeroportos brasileiros em meados dos anos 2010. O objetivo foi investigar se elementos físicos da configuração interna das aeronaves e indicadores de densidade estão associados a variações no preço, controlando fatores operacionais, concorrenciais, de rede e de perfil do passageiro. Sabe-se que, atualmente, esses atributos a bordo são amplamente utilizados pelas companhias aéreas para diferenciar produtos e serviços, de forma a conceder benefícios a clientes premium e alinhar a qualidade intrínseca do assento à disposição a pagar do passageiro. Quanto mais conciliadas forem essas dimensões do esforço de vendas das firmas, mais eficiente e competitiva ela se torna no mercado. Ao examinar esse tema em um período no qual tais práticas ainda eram incipientes, torna-se possível identificar potenciais caminhos de aprimoramento e compreender de forma mais precisa como a alocação de assentos vinha sendo administrada pelas empresas e evoluiu desde então.

A estratégia de estimação empregou o método PDS-LASSO, que realiza seleção de variáveis em bases de dados com grande número de covariadas para modelar fenômenos complexos. No presente estudo, o conjunto de controles considerado incluía efeitos de data da pesquisa, horário do voo, aeroportos de origem e destino e perfil de passageiro. Esses blocos foram submetidos ao procedimento de seleção, que extrai subconjuntos relevantes para cada especificação, considerando a relação estatística dessas covariadas com o preço e com as variáveis de interesse.

Alguns dos principais resultados foram os seguintes. Há evidências de associação negativa entre densidade de assentos e preços, sugerindo que aeronaves com maior número de fileiras por unidade de cabine tendem a apresentar tarifas menores, compatíveis com efeitos de diluição de custos. Esse ponto é relevante no contexto da proposta do assento Skyrider (fabricante Aviointeriors), discutida ao longo do trabalho, que se configura em um conceito de assentos em posição quase vertical, em formato de selim de bicicleta, permitindo aumentos substanciais no número de passageiros a bordo. Outros conceitos também podem produzir efeitos semelhantes de elevação significativa da densidade, como o Air Lair com casulos de dois andares (Factorydesign), o Zephyr Seat com nível superior adicional (Zephyr Aerospace) e o Cabin Hexagon com geometria otimizada para acréscimo de lugares (Zodiac Seats). Os resultados obtidos no presente trabalho sinalizam que aumentos de densidade são estatisticamente associados a preços mais baixos, o que sugere que inovações em layouts que amplifiquem a capacidade, como esses e outros projetos de assentos e layout de cabine em desenvolvimento, poderiam beneficiar as companhias aéreas, e em particular as empresas voltadas à atração de passageiros com alta elasticidade-preço da demanda. Contudo, como esses indicadores captam efeitos estruturais e não preferências reveladas do passageiro, a viabilidade econômica de propostas como o Skyrider depende de quanto os consumidores estariam dispostos a aceitar reduções de conforto em troca de tarifas menores, algo que os dados deste período não permitem avaliar diretamente.

Adicionalmente, observou-se que ocupantes de assentos do meio pagavam, ceteris paribus, preços de passagens mais altos que passageiros de corredor ou janela, um padrão que as interações com antecedência de compra indicaram estar associado à dinâmica de compra tardia, e não a qualquer valorização específica desse tipo de assento. Essa evidência é coerente com o fato de que, em ambientes comerciais sem cobrança explícita pela marcação do assento, posições a bordo mais desejáveis tendem a não estar tão disponíveis quando os preços já se encontram elevados pelos mecanismos de gerenciamento de receita. Essa ineficiência alocativa ajuda a explicar por que a tarifação de seleção de assentos se torna um instrumento importante de segmentação de mercado, pois permite que as empresas organizem a distribuição dos assentos de forma mais alinhada à disposição a pagar dos passageiros. O trabalho portanto contribui ao mostrar uma base econômica que ajuda a explicar a racionalidade de uma das principais fontes de receitas auxiliares das companhias aéreas.

Por fim, outras variáveis locacionais, como assentos com espaço extra, assentos de emergência e última fileira, não apresentaram significância estatística, indicando um potencial para investimento em práticas de segmentação mais eficazes. Todos esses resultados são compatíveis com estratégias que as companhias aéreas passaram a desenvolver intensamente, nas quais as estratégias de segmentação de clientes se tornou um dos avanços mais marcantes da aviação comercial recente. O uso crescente de atributos a bordo, incluindo a tarifação dinâmica da seleção de assentos e aprimoramentos na diferenciação de conforto, aponta para a continuidade desse processo e sugere que práticas similares podem seguir se expandindo nos próximos anos, acompanhando a evolução tecnológica e comercial dos layouts de cabine.

## REFERÊNCIAS